\documentclass[11pt]{article}
\usepackage{amssymb}
\usepackage{epsfig}
\newlength{\bredde}
\def\slash#1{\settowidth{\bredde}{$#1$}\ifmmode\,\raisebox{.15ex}{/}
\hspace*{-\bredde} #1\else$\,\raisebox{.15ex}{/}\hspace*{-\bredde} #1$\fi}
\newcommand{\be}{\begin{equation}}
\newcommand{\ee}{\end{equation}}
\newcommand{\bea}{\begin{eqnarray}}
\newcommand{\eea}{\end{eqnarray}}
\newcommand{\nn}{\nonumber}
\newcommand{\al}{\alpha}

\newcommand{\Lh}{\hat{L}^{\nu}}
\newcommand{\one}{\mbox{\bf 1}}

\def\Sl#1{\rlap{\raisebox{.15ex}{$\mskip 4 mu /$}}#1}  

\newcommand{\sect}[1]{\setcounter{equation}{0}\section{#1}}

\def\hx{\hat{x}}
\def\hy{\hat{y}}
\def\hm{\hat{m}}
\def\hs{\hat{s}}
\def\hmu{\hat{\mu}}
\def\hd{\hat{\delta}}

\begin{document}
\topmargin -1.4cm
\oddsidemargin -0.8cm
\evensidemargin -0.8cm
\title{\Large{{\bf 
Individual Eigenvalue Distributions of Chiral Random Two-Matrix Theory
and the Determination of $F_\pi$
}}}

\vspace{1.5cm}
\author{~\\{\sc G.~Akemann}$^1$ and {\sc P.~H.~Damgaard}$^{2}$
\\~\\$^1$Department of Mathematical Sciences \& BURSt Research Centre\\
Brunel University West London\\ 
Uxbridge UB8 3PH\\United Kingdom
\\~\\
$^2$The Niels Bohr Institute\\The Niels Bohr International Academy\\
Blegdamsvej 17\\ DK-2100 Copenhagen\\ Denmark}

\date{}
\maketitle
\vfill
\begin{abstract}
Dirac
operator eigenvalues split into two when subjected to two different 
external vector sources. In a specific finite-volume scaling regime
of gauge theories with fermions,
this problem can be mapped to a chiral Random Two-Matrix Theory. 
We derive analytical expressions to leading order in the associated
finite-volume expansion, showing how
individual Dirac eigenvalue distributions and their correlations
equivalently
can be computed directly from the effective chiral Lagrangian in 
the epsilon-regime. Because of its equivalence to 
chiral Random Two-Matrix Theory, we use 
the latter for all explicit computations.  
On the mathematical side, we define and determine gap
probabilities and individual eigenvalue distributions in that
theory at finite $N$, and also derive the relevant scaling limit
as $N$ is taken to infinity. 
In particular, the gap probability for one Dirac eigenvalue is given in terms
of a new kernel that depends on the external vector source. This
expression may give a new and simple way of determining the pion decay
constant $F_\pi$ from lattice gauge theory simulations.

\end{abstract}
\vfill

\thispagestyle{empty}
\newpage

\renewcommand{\thefootnote}{\arabic{footnote}}
\setcounter{footnote}{0}

\sect{Introduction}\label{defs}

One of the most challenging -- and perhaps most interesting -- 
problems associated
with lattice gauge theory simulations of QCD is that of the chiral limit.
Based on a variety of different approaches it is now possible to perform
numerical simulations of the theory with two light dynamical quark flavours, 
at least in modest space-time volumes. By ``light'' quarks we mean
quarks that are very close to the actual physical masses of the
$u$ and $d$ quarks in QCD. Even if the masses of the physical $u$ 
and $d$ quarks
had turned out to be much heavier (on the typical QCD scale $\Lambda_{QCD}$),
one would like to explore the chiral limit of the theory in it own right.
This is because the theory in this limit separates into to two disjoint
regimes, of which the low-energy part can be treated in a systematic
manner by means of effective field theory: the chiral Lagrangian based
on the spontaneous breaking of chiral symmetry. This low-energy theory
of QCD with very light quarks can be understood in much the same way
that the low-energy limit of QCD without quarks matches on to an effective
string theory description, and both limits are of interest.

The so-called $\epsilon$-regime of QCD \cite{LS} is particularly useful
for studying the chiral limit of QCD in finite volume. It is well known
how a universality class of chiral Random Matrix Theory \cite{Jacetal} 
provides 
an intriguing alternative description of the leading-order expressions 
for Dirac
operator eigenvalue correlation functions in this regime, results
that also can be derived directly from the low-energy effective field
theory \cite{DOTV,BA}. Even the distributions of individual Dirac operator
eigenvalues follow from a systematic expansion in the chiral
Lagrangian framework \cite{AD}. All of these analytical results depend
on just one single low-energy constant of QCD, that of the 
infinite-volume chiral condensate
$\Sigma$. From a lattice gauge theory viewpoint, this provides a new
and unusual way of determining this low-energy constant of QCD 
by measuring the lowest-lying Dirac operator eigenvalues. For some
numerical analyses see, $e.g.$, refs. \cite{Sigmasim0,Sigmasim}.

Recently, a new scheme was proposed which uses Dirac operator eigenvalues 
for determining the pion decay constant $F_{\pi}$ in a somewhat similar manner
\cite{Imiso}. Based on the chiral Lagrangian formulation, the
suggested method made use of
a spectral 2-point function associated with two different
Dirac operators,
\bea
D_1\psi_1^{(n)} &\equiv & [\Sl{D}(A)+i\mu_1\gamma_0]\psi_1^{(n)} ~=~
 i\lambda_1^{(n)}\psi_1^{(n)} \ , \cr
D_2\psi_2^{(n)} &\equiv & [\Sl{D}(A)+i\mu_2\gamma_0]\psi_2^{(n)} ~=~
 i\lambda_2^{(n)}\psi_2^{(n)}\ ,
\label{Ddef}
\eea
corresponding, in the case $\mu \equiv \mu_1 = -\mu_2$,  
to {\em imaginary isospin chemical potential}. Equivalently,
the two Dirac operators  (\ref{Ddef}) are simply in a constant background
Abelian gauge field, but with different ``charges''. In the $\epsilon$-regime,
the chemical potential $\mu$ couples directly to $F_{\pi}$ in the
form of the finite-volume scaling variable $\hat{\mu} = \mu F_{\pi}\sqrt{V}$.
Because the sensitivity to $\mu$ is quite drastic for the spectral
2-point function, this provides a clean method for extracting $F_{\pi}$.
There is sensitivity to the parameter $\hat{\mu}$ (and hence $F_{\pi}$)
also in other observables in the $\epsilon$-regime \cite{MT,Luz}. 
Alternatively, one may use a real chemical potential to determine $F_\pi$ 
\cite{AW}, with the same finite-volume scaling. The Dirac spectrum
is complex in that case.

The chiral Lagrangian approach of ref. \cite{Imiso} can, to leading order
in the $\epsilon$-regime, also be re-cast in terms of Random Matrix Theory,
this time a Random Two-Matrix Theory \cite{ADOS}. All eigenvalue
density correlations are equivalent in the two theories \cite{Imiso,BA}. 
One loop corrections
to both $\Sigma$ \cite{LS} and $F_{\pi}$ \cite{Tom} have been computed in 
the $\epsilon$-expansion. To that order they simply amount to
finite-volume corrections to the infinite-volume quantities
$\Sigma$ and $F_{\pi}$; the effective theory otherwise remains
unchanged. It is of course important to know the size of these
finite-size corrections if one wishes to determine $\Sigma$ and $F_{\pi}$
from the eigenvalues of the Dirac operator by means of lattice
gauge theory simulations at finite volume. An alternative method
for extracting $\Sigma$ and $F_{\pi}$ in the $\epsilon$-regime of
QCD can be based on fits to vector and axial vector two-point
correlations functions; also here finite-volume corrections are
know analytically at sectors of fixed gauge field topology 
\cite{VA}. See the recent review \cite{Necco} for a summary
of these different approaches.

In the Random Matrix Theory
formulation analytical computations are substantially simplified, and in
ref. \cite{ADOS} 
all possible spectral density correlation functions
associated with the two Dirac operators were found analytically. This
includes all spectral functions in both the quenched and unquenched theory,
and even all spectral correlation functions associated with ``partially
quenched'' spectral correlation functions, where there is no back-reaction
of the chemical potential on the gauge field configurations. This latter
set of spectral correlation functions give the most fruitful way of extracting
$F_{\pi}$ from lattice data since one can make use of ordinary configurations
without chemical potential. Once all spectral correlation functions
are known, one should in principle have all spectral data, and thus be able
to reconstruct {\em individual eigenvalue distributions} as well. Indeed,
this was precisely what was found in ref. \cite{AD} for the case without
imaginary chemical potential. In the first part of this paper we show that
this is also the case here.\footnote{A preliminary account of 
this was presented
at a conference last year \cite{AD07}. A comparison with Monte Carlo data
from lattice gauge theory was presented at the same meeting
\cite{Tom0}.} It was also shown in \cite{A07} that
in the quenched theory in the limit of large chemical potential
all spectral and individual eigenvalue correlations factorise 
into quenched one-matrix theory quantities.

In the second part of this paper we aim at determining in as closed form
as possible the precise analytical expressions for individual eigenvalues
distributions. Such expressions may turn out to be very useful alternatives
for extracting $F_{\pi}$ from lattice simulations in those cases where only
a very small number of ``good'' eigenvalues are available, and where 
it therefore may be difficult to construct the spectral 2-point correlation 
function with good statistics. Here we concentrate on the smallest eigenvalue,
but provide the general framework for computing others. Based on the
Random Two-Matrix Theory representation, we derive an explicit and quite 
compact representation for any finite $N$. Taking the scaling limit
with $N \to \infty$,
this provides the sought-for analytical expression for the lowest
Dirac operator eigenvalue distribution in the appropriate finite-volume
scaling regime. Remarkably, the final expression is not much more
involved than the one without external vector sources.

From the point of view of mathematical physics, the resulting solution
for the distribution of the smallest eigenvalue in the chiral Random
Two-Matrix Theory is of interest in its own right. The solution cannot
be mapped on to an analogous one-matrix theory, so the distribution
is new and presumably corresponds to a new universality class that is
parametrised by one real number $\hat{\delta}$. For this reason we include
a rather detailed derivation, even though the resulting formula
is all that is needed for the purpose of applications to lattice
gauge theory simulations.  

\sect{Eigenvalue correlations in chiral Random Two-Matrix Theory
}

We start by giving the theory we will solve for individual eigenvalue
correlations, chiral perturbation theory in the epsilon regime with imaginary
chemical potential
\be
{\cal Z}_{\nu}^{(N_f)} =
\int_{U(N_f)} dU \,(\det U)^{\nu}e^{\frac{1}{4}VF_\pi^2
{\rm Tr} [U,B][U^\dagger,B] + \frac{1}{2} \Sigma V
{\rm Tr}({\cal M}^\dagger U + {\cal M}U^\dagger)} ~.
\label{ZXPT}
\ee
Here the matrix $B=$diag$(\mu_1\one_{N_1},\mu_2\one_{N_2})$ contains the two 
different
chemical potentials, and 
${\cal M}=\mbox{diag}(m_1,\ldots,m_{N_f})$ is the quark mass matrix of the 
$N_1+N_2=N_f$ flavours.

This theory and all its spectral density correlation functions are completely
equivalent to  the chiral Two-Matrix Theory with imaginary chemical 
potential that was introduced in ref. \cite{ADOS}. The equivalence for the 
two-point function follows from \cite{Imiso}, for all higher density
correlations it was proven in \cite{BA}. It is defined as 
\bea
{\cal Z}_{\nu}^{(N_f)}&\sim&
 \int d\Phi  d\Psi~ 
e^{-{N}{\rm Tr}\left(\Phi^{\dagger}
\Phi + \Psi^{\dagger}\Psi\right)}
\prod_{f1=1}^{N_1} \det[{\cal D}_1 + m_{f1}] 
\prod_{f2=1}^{N_2} \det[{\cal D}_2 + m_{f2}] \ ,
\label{ZNf}
\eea
where ${\cal D}_f$ is given by
\bea
{\mathcal D}_f = \left( \begin{array}{cc}
0 & i \Phi + i \mu_f \Psi \\
i \Phi^{\dagger} + i \mu_f \Psi^{\dagger} & 0
\end{array} \right) \ \ ,\ \ f=1,2\ ~.
\eea
The operator remains 
anti-Hermitian because the chemical potentials are imaginary. 
Both $\Phi$ and $\Psi$ are complex rectangular 
matrices of size $N\times (N+\nu)$, where both $N$ and $\nu$ are integers.
The index $\nu$ corresponds to gauge field topology in the usual way.

Referring to ref. \cite{ADOS} for details, we immediately write down
the corresponding eigenvalue representation:
\bea
{\cal Z}_{\nu}^{(N_f)}
&=& \int_0^{\infty} \prod_{i=1}^N\left(dx_idy_i (x_iy_i)^{\nu+1}
\prod_{f1=1}^{N_1} (x_i^2+m_{f1}^2)
\prod_{f2=1}^{N_2} (y_i^2+m_{f2}^2) \right) \cr
&\times&\Delta_N(\{x^2\})\Delta_N(\{y^2\})\det\left[I_{\nu}(2 d N x_i y_j)
\right] 
e^{-N \sum_i^N c_1 x_i^2 + c_2 y_i^2 }, \label{evrep}
\eea
up to an irrelevant normalisation factor.
Here the $x_i$'s and $y_i$'s are the real and non-negative entries in the
diagonal matrices $X$ and $Y$, defined by
\bea
\Phi_1 &\equiv& \Phi + \mu_1 \Psi\ =\ U_1XV_{1}^{\dagger} \ ,\cr
\Phi_2 &\equiv& \Phi + \mu_2 \Psi\ =\ U_2YV_{2}^{\dagger} ~.
\eea
Because of this redefinition the matrices $\Phi_i$ become coupled in the
exponent, leading to the above structure after integration out the unitary
matrices $U_i$ and $V_i$. This leads to 
he following combinations of the two chemical potentials in eq. (\ref{evrep}):
\bea
c_1 &=& (1+\mu_2^2)/\delta^2  \ ,\ \ \ \ c_2 \ =\ (1+\mu_1^2)/\delta^2 \ ,\nn\\
d &=& (1+\mu_1\mu_2)/\delta^2 \ ,\ \  1-\tau \ =\ d^2/(c_1c_2)\ ,\nn\\
\delta &=& \mu_2 - \mu_1\ ,
\label{cdtdef}
\eea
where the latter will become useful in section \ref{exactpE}.

The joint probability distribution function which is proportional to the 
integrand in eq. (\ref{evrep}) is defined as 
\bea
{\cal P}_{\nu}^{(N_f)}(\{x\},\{y\};\{m_1\},\{m_2\}) 
&=& \frac{1}{{\cal Z}_{\nu}^{(N_f)}}\prod_{i=1}^N\left((x_iy_i)^{\nu+1}
\prod_{f1=1}^{N_1} (x_i^2+m_{f1}^2)
\prod_{f2=1}^{N_2} (y_i^2+m_{f2}^2) \right) \cr
&\times&\Delta_N(\{x^2\})\Delta_N(\{y^2\})\det\left[I_{\nu}(2 d N x_i
  y_j)\right]  e^{-N \sum_i^N c_1 x_i^2 + c_2 y_i^2 } , \label{jpdf}
\eea
where $\Delta_N(\{x^2\})=\prod_{j>i}^N(x_j^2-x_i^2)$ 
is the Vandermonde determinant. 
It is normalised to unity
\be
1 \ =\ 
\int_0^{\infty} \prod_{i=1}^Ndx_idy_i \ 
{\cal P}_{\nu}^{(N_f)}(\{x\},\{y\};\{m_1\},\{m_2\}) \ .
\label{jpdfnorm}
\ee
From the joint probability distribution we can define an 
{\em $(n,k)$-density correlation function}
\bea
R_{k,l}(\{x\}_k,\{y\}_l) &\equiv&
\frac{N!^2}{(N-k)!(N-l)!}
\int_0^{\infty}  \prod_{i=k+1}^N dx_i \prod_{j=l+1}^N dy_j 
\ {\cal P}_{\nu}^{(N_f)}(\{x\},\{y\};\{m_1\},\{m_2\}) \nn\\
&=&
\frac{N!^2}{(N-k)!(N-l)!}
\frac{1}{{\cal Z}_{\nu}^{(N_f)}} 
\\
&\times&
\int_0^{\infty}  \prod_{i=k+1}^N dx_i \prod_{j=l+1}^N dy_j 
\det\left[w_\nu^{(N_f)}(x_i,y_j)\right]
\Delta_N(\{x^2\})\Delta_N(\{y^2\}), \nn
\label{Revrep}
\eea
where we have moved the exponential and masses 
into the determinant, introducing
\be
w_\nu^{(N_f)}(x_i,y_j)\ \equiv\ 
(x_i y_j)^{\nu+1}\ e^{-N(c_1 x_i^2 + c_2 y_j^2) } 
I_{\nu}(2 d N x_i  y_j)
\prod_{f1=1}^{N_1} (x_i^2+m_{f1}^2)
\prod_{f2=1}^{N_2} (y_j^2+m_{f2}^2)
\ .
\label{weightdef}
\ee
Obviously $R_{0,0}=1$ is normalised to unity.
The $R_{k,l}(\{x\}_k,\{y\}_l)$
can be expressed in terms of a determinant of four different kernels. 
These are given by the (bi-)orthogonal polynomials and their 
integral transforms
with respect to the weight function eq. (\ref{weightdef}), and we refer to 
\cite{ADOS} for details.
The $R_{k,l}(\{x\}_k,\{y\}_l)$
will be the building blocks to compute the gap probabilities as well as
the distributions of individual eigenvalues of both type $x$ and $y$.

We define the following gap probabilities as  
\bea
E_{k,l}(s,t) &\equiv& \frac{N!^2}{(N-k)!(N-l!)}
\int_0^s dx_1\ldots dx_{k}\int_s^\infty dx_{k+1}\ldots dx_N 
\int_0^t dy_1\ldots dy_{l}\int_t^\infty dy_{l+1}\ldots dy_N 
\nn\\
&&\times\ {\cal P}_{\nu}^{(N_f)}(\{x\},\{y\};\{m_1\},\{m_2\})
\ , \ \ \mbox{for}\ k,l=0,1,\ldots,N\ \ ,
\label{Ekl}
\eea
where in the sequel we suppress the dependence on masses and topology  for
simplicity. 
The $E_{k,l}(s,t)$ 
give the probability for general $k,l\in\ \{0,\ldots,N\}$ that the interval 
$[0,s]$ is occupied by $k$ $x$-eigenvalues of ${\cal D}_1$
and $[s,\infty)$ is 
occupied by $(N-k)$ $x$-eigenvalues, and that the interval 
$[0,t]$ is occupied by $l$ $y$-eigenvalues of ${\cal D}_2$
and $[t,\infty)$ is 
occupied by $(N-l)$ $y$-eigenvalues. It also depends on the masses and on 
$\mu_{1,2}$ which we have suppressed here. 

Similarly we can define the probability to find the $k$-th $x$-eigenvalue 
at value $x_k=s$, and the $l$-th $y$-eigenvalue 
at value $y_l=t$, to be 
\bea
p_{k,l}(s,t) &\equiv& k {N \choose k}l {N \choose l}
\int_0^s dx_1\ldots dx_{k-1}\int_s^\infty dx_{k+1}\ldots dx_N 
\int_0^t dy_1\ldots dy_{l-1}\int_t^\infty dy_{l+1}\ldots dy_N 
\nn\\
&\times&  
{\cal P}_{\nu}^{(N_f)}(x_1,\ldots,x_{k-1},x_k=s,x_{k+1},\ldots, x_N,
y_1,\ldots,y_{l-1},y_l=t,y_{l+1},\ldots, y_N;\{m_1\},\{m_2\}).\nn\\
\label{pkl}
\eea
Here the eigenvalues are ordered, $x_1\leq\ldots\leq x_N$ and
$y_1\leq\ldots\leq y_N$, and obviously $k,l\geq1$.
The fact that the $p_{k,l}(s,t)$ are probabilities that are normalised as
\be
\int_0^\infty ds\int_0^\infty dt\ p_{k,l}(s,t)\ =\ 1\ \ ,
\ee
can be seen along the same lines as for a single set of Dirac operator
eigenvalues, as was shown in the appendix of \cite{AD}.

Because we have two sets of eigenvalues we may also define mixed
gap-probability distributions as they will occur in intermediate steps. There
are two different functions defined as
\bea
Ep_{k,l}(s,t) &\equiv& \frac{N!}{(N-k)!}l {N \choose l}
\int_0^s dx_1\ldots dx_{k}\int_s^\infty dx_{k+1}\ldots dx_N 
\int_0^t dy_1\ldots dy_{l-1}\int_t^\infty dy_{l+1}\ldots dy_N 
\nn\\
&&\times\ {\cal P}_{\nu}^{(N_f)}(\{x\},
y_1,\ldots,y_{l-1},y_l=t,y_{l+1},\ldots, y_N;\{m_1\},\{m_2\})
\ , \nn\\
&&\mbox{for}\ k=0,\ldots,N\ \ \mbox{and} \ l=1,\ldots,N\ \ ,
\label{Epkl}\\
&&\ \nn\\
pE_{k,l}(s,t) &\equiv& k {N \choose k}\frac{N!}{(N-l!)}
\int_0^s dx_1\ldots dx_{k-1}\int_s^\infty dx_{k+1}\ldots dx_N 
\int_0^t dy_1\ldots dy_{l}\int_t^\infty dy_{l+1}\ldots dy_N 
\nn\\
&&\times\ {\cal P}_{\nu}^{(N_f)}(x_1,\ldots,x_{k-1},x_k=s,x_{k+1},\ldots, x_N,
\{y\};\{m_1\},\{m_2\})
\ , \nn\\
&&\mbox{for}\ k=1,\ldots,N \ \ \mbox{and} \ l=0,\ldots,N\ \ .
\label{pEkl}
\eea 
The first quantity eq. (\ref{Epkl}) gives the probability that 
$[0,s]$ is occupied by $k$ of the $x$-eigenvalues of ${\cal D}_1$
and $[s,\infty)$ is occupied by $(N-k)$ of the $x$-eigenvalues,
given that $y_l=t$, where the $y$-eigenvalues are ordered.
The second quantity eq. (\ref{pEkl}) gives the probability that 
$[0,t]$ is occupied by $l$ of the $y$-eigenvalues of ${\cal D}_2$
and $[t,\infty)$ is occupied by $(N-l)$ of the $y$-eigenvalues,
given that $x_k=s$, where again the $x$-eigenvalues are ordered.

These definitions include for example the probability
$pE_{k,0}(s,t=0)$ to find an eigenvalue of the first type at $x=s$, where all
$y$-eigenvalues are integrated out. We will return to this in section 
\ref{exactpE}.  

\sect{Gap probabilities and individual eigenvalues from densities}

We use the simple identity 
\be
(a-b)^j \ =\ \sum_{l=0}^j (-1)^l {j \choose l}
a^{j-l} b^l\ ,
\label{ab}
\ee
and choose $a=\int_0^\infty dx$ and $b=\int_0^s dx$ to replace all 
the $(N-k)$ $dx$-integrals $\int_s^\infty dx$ in eq. (\ref{Ekl}) 
by $a-b$, and likewise for the corresponding $y$-integrations. We obtain
\bea
E_{k,l}(s,t) &=& \frac{N!^2}{(N-k)!(N-l)!}
\int_0^s dx_1\ldots dx_{k} \int_0^t dy_1\ldots dy_{l} \nn\\
&&\times\sum_{i=0}^{N-k} (-1)^i {N-k \choose i}
\left(\int_0^\infty\right)^{N-k-i}\left(\int_0^s\right)^i dx_{k+1}\ldots dx_N
\nn\\
&&\times\sum_{j=0}^{N-l} (-1)^j {N-l \choose j}
\left(\int_0^\infty\right)^{N-l-j}\left(\int_0^t\right)^j dy_{l+1}\ldots dy_N 
\ {\cal P}_{\nu}^{(N_f)}(\{x\},\{y\};\{m_1\},\{m_2\})
\nn\\
&=& \sum_{i=0}^{N-k}\sum_{j=0}^{N-l}  \frac{(-1)^{i+j}}{i!j!} 
\int_0^s  dx_1\ldots dx_{k+i}\ \int_0^t  dy_1\ldots dy_{l+j}\ 
R_{k+i,l+j}(x_1,\ldots,x_{k+i},\ y_1,\ldots,y_{l+j}).
\nn\\
\label{EklR}
\eea
Here we have used the invariance of the joint probability distribution under 
permutations of both $\{x\}$ and $\{y\}$. 
The formula (\ref{EklR}) neatly expresses the gap probability
in terms of spectral correlation functions of both sets of eigenvalues.
The latter can be derived from $k$-point resolvents in chiral perturbation
theory by enlarging eq. (\ref{ZXPT}) to the corresponding supergroup integral,
see \cite{BA}. 
Thus we have shown how in this setting also gap probabilities and
individual eigenvalue distributions follow at the level of the chiral 
Lagrangian.

We may introduce a generating functional for all gap probabilities,  
\be
E(s,t;\xi,\eta)\ \equiv\ \sum_{i,j=0}^{N} (-\xi)^i(-\eta)^j \frac{1}{i!j!} 
\int_0^s  dx_1\ldots dx_{i}\ \int_0^t  dx_1\ldots dx_{j}\ 
R_{i,j}(x_1,\ldots,x_{i},\ y_1,\ldots,y_{j})
\ ,
\label{Egen}
\ee
where the term at $i=j=0$ gives unity.
It immediately follows that 
\be
E_{k,l}(s,t)\ =\ (-1)^{k+l} \left.
\frac{\partial^{k}}{\partial\xi^{k}} \ 
\frac{\partial^{l}}{\partial\eta^{l}} \ 
E(s,t;\xi,\eta) \right|_{\xi=1,\eta=1} \ , \ \ \mbox{for}\ k,l=0,1,
\ldots,N\ \ .
\label{Ekpartial}
\ee
We will now relate gap probabilities, mixed and individual eigenvalue
distributions to density correlations. It can be easily shown that 
\be
 \frac{\partial}{\partial s} E_{k,l}(s,t)
\ =\ k!\left( pE_{k,l}(s,t) - 
pE_{k+1,l}(s,t)\right)\ .
\label{EklpEkl}
\ee
For $k=l=0$ we have 
\be
 \frac{\partial}{\partial s} E_{0,0}(s,t)
\ =\ - pE_{1,0}(s,t)\ ,
\ee
as from the definition $pE_{k,l}$ has $k\geq 1$,
and thus we set $pE_{0,l}(s,t)\equiv0$. Similarly it follows
\be
 \frac{\partial}{\partial t} E_{k,l}(s,t)
\ =\ l!\left( Ep_{k,l}(s,t) - 
Ep_{k,l+1}(s,t)\right)\ ,
\label{EklEpkl}
\ee
where again  $Ep_{k,0}(s,t)\equiv0$. 
If we differentiate the mixed correlators we obtain
\bea
 \frac{\partial}{\partial s} Ep_{k,l}(s,t)
&=& k!\left( p_{k,l}(s,t) - 
p_{k+1,l}(s,t)\right)\ ,
\label{Epklpkl}\\
 \frac{\partial}{\partial t} pE_{k,l}(s,t)
&=& l!\left( p_{k,l}(s,t) - 
p_{k,l+1}(s,t)\right)\ .
\label{pEklpkl}
\eea
Finally, if we differentiate the gap probabilities twice we arrive at 
\be
 \frac{\partial^2}{\partial s\partial t} E_{k,l}(s,t)
\ =\ k!\ l!\left( p_{k,l}(s,t) - p_{k+1,l}(s,t)
-p_{k,l+1}(s,t) + p_{k+1,l+1}(s,t)\right)\ .
\label{Eklpkl}
\ee
Of course the order of differentiation does not matter, as one can easily
convince oneself. 
The boundary conditions to be imposed here and in eqs. (\ref{Epklpkl}) and 
(\ref{pEklpkl}) follow from 
\bea
 \frac{\partial^2}{\partial s\partial t} E_{k,0}(s,t)
&=&-\ k!\left( p_{k,1}(s,t) - p_{k+1,1}(s,t)\right)\ ,\nn\\
 \frac{\partial^2}{\partial s\partial t} E_{0,l}(s,t)
&=&-\ l!\left( p_{1,l}(s,t) - p_{1,l+1}(s,t)\right)\ .
\label{Eklpklbc}
\eea
Again from the definitions we have 
$p_{k,0}(s,t)=p_{0,l}(s,t)\equiv0$. 

The probabilities $p_{k,l}(s,t)$ can be solved for the (mixed) gap
probabilities in three different ways. Summing over $k$ in
eq. (\ref{Epklpkl}),  
or over $l$ in eq. (\ref{pEklpkl})
we obtain 
\bea
p_{n+1,l}(s,t) &=& - \sum_{k=0}^n \frac{1}{k!} 
\frac{\partial}{\partial s} Ep_{k,l}(s,t)\ ,\nn\\
p_{k,n+1}(s,t) &=& - \sum_{l=0}^n \frac{1}{l!} 
\frac{\partial}{\partial t} pE_{k,l}(s,t) \ .
\label{pklI}
\eea
Alternatively one can sum over both $k$ and $l$ in eq. (\ref{Eklpkl}) to
obtain an expression in terms of gap probabilities alone
\be
p_{n+1,q+1}(s,t)\ =\ +\sum_{k=0}^n \sum_{l=0}^q \frac{1}{k!\,l!} 
\frac{\partial^2}{\partial s\partial t} E_{k,l}(s,t)\ .
\label{pklII}
\ee

Let us give some examples. 
For the simplest case of $k=l=0$ we get the probability that the interval
$[0,s]$ is free of $x$-, and the interval $[0,t]$ free of $y$-eigenvalues:
\be
E_{0,0}(s,t)\ =\ 
\int_s^{\infty} \prod_{i=1}^N dx_i
\int_t^{\infty} \prod_{j=1}^N dy_j  
\ {\cal P}_{\nu}^{(N_f)}(\{x\},\{y\};\{m_1\},\{m_2\}) \ ,
\ee
and we obtain
\be
\frac{\partial^2}{\partial s\partial t} E_{0,0}(s,t)\ =\ p_{1,1}(s,t)\ .
\label{E00}
\ee
Explicitly we have for this gap probability the expansion eq. (\ref{EklR})
given already in \cite{AD07}
\bea
E_{0,0}(s,t) &=&
1-\int_0^s dx\, R_{1,0}(x)- \int_0^t dy\, R_{0,1}(y)
+ \frac12 \int_0^s dx_1dx_2\, R_{2,0}(x_1,x_2)
\nn\\
&&+ \frac12 \int_0^t dy_1dy_2\, R_{0,2}(y_1,y_2)\ +\ \ldots \nn\\
&&+ \int_0^s dx\int_0^t dy  R_{1,1}(x,y)
- \frac12 \int_0^s dx_1dx_2 \int_0^t dy\, R_{2,1}(x_1,x_2,y)
\nn\\
&&- \frac12 \int_0^s dx\int_0^t dy_1dy_2\, R_{1,2}(x,y_1,y_2) \ +\ \ldots
\ \ .\label{E00ex} 
\eea
The terms in the first two lines containing only $s$- or $t$-dependent 
integrals are 
annihilated by the differentiation in eq. (\ref{E00}),
and we get to the same order 
\be
p_{1,1}(s,t)\ =\ R_{1,1}(s,t)\ -\  \int_0^s dx\, R_{2,1}(x,s,t)
\ -\ \int_0^t dy\, R_{1,2}(s,t,y)\ +\ \ldots\ .
\label{p00ex}
\ee

\sect{An exact expression for the first eigenvalue distribution}
\label{exactpE}

In this section we derive a closed expression for an individual
eigenvalue distribution, the probability to find the first eigenvalue of  
${\cal D}_1$ at $s$ irrespective of the position of the 
${\cal D}_2$-eigenvalues. Our solution given in terms of a new kernel and
polynomials holds for any number of flavours $N_1$ and $N_2$ and arbitrary
chemical potentials $\mu_1$ and $\mu_2$. In particular, we can
partially quench the type-1 flavours (putting $N_1=0$) 
with $\mu_1\neq0$ in gauge theory with type-2 physical sea-quark
flavours ($i.e., N_2\neq0$) with
$\mu_2=0$. This case is probably
the most interesting for applications to lattice QCD.

\subsection{The finite-$N$ solution}

We first consider the gap probability that the interval 
$[0,s]$ is empty of $x$-eigenvalues,
\be
E_{0,0}(s,t=0) =
\int_s^\infty dx_1\ldots dx_N 
\int_0^\infty dy_1 \ldots dy_N \ 
{\cal P}_{\nu}^{(N_f)}(\{x\},\{y\};\{m_1\},\{m_2\})\ .
\label{1gapdef}
\ee
From this the sought probability follows by differentiation,
$pE_{1,0}(s,t=0)=-\partial_s E_{0,0}(s,t=0)$. 
We will now 
perform a series of steps before arriving at an exact expression for
finite $N$. The appropriate 
large-$N$ scaling limit will be taken in the next subsection.

First, recalling the definition of the joint probability distribution 
eq. (\ref{jpdf}) we can use the fact that
all $y$-eigenvalues are integrated out in eq. (\ref{1gapdef})
for a symmetry argument. The Vandermonde
determinant $\Delta_N(\{y^2\})$ and the determinant of the Bessel
function are antisymmetric. Therefore we can replace the latter by its diagonal
part times $N!$
\bea
E_{0,0}(s,0) &=&\frac{N!}{{\cal Z}_\nu^{(N_f)}}
\int_s^\infty dx_1\ldots dx_N 
\int_0^\infty dy_1 \ldots dy_N 
\Delta_N(\{x^2\})\Delta_N(\{y^2\}) 
 \cr
&\times&\prod_{i=1}^N\left((x_iy_i)^{\nu+1}\  e^{-N (c_1 x_i^2 + c_2 y_i^2) }
I_{\nu}(2 d N x_i y_i)
\prod_{f1=1}^{N_1} (x_i^2+m_{f1}^2)
\prod_{f2=1}^{N_2} (y_i^2+m_{f2}^2) 
\right).
\label{E1}
\eea
In the next step we use a known identity relating 
the Laguerre weight times the $I$-Bessel function to  
an {\it infinite} sum over Laguerre polynomials 
(see e.g. eq. (B.7) in \cite{ADOS}). With this decomposition we
can exploit the orthogonality properties of these polynomials. 
For simplicity of the proof 
we will first consider $N_1=N_2=1$ $(N_f=2)$ with masses $m_1$ and
$m_2$,  and later give the general result for
any numbers of flavours. We thus have 
\bea
E_{0,0}(s,0) &=& \frac{N!}{{\cal Z}_\nu^{(1+1)}}
\int_s^\infty dx_1\ldots dx_N 
\int_0^\infty dy_1 \ldots dy_N 
\Delta_N(\{x^2\})\Delta_N(\{y^2\}) \prod_{i=1}^N (x_i^2+m_{1}^2)
(y_i^2+m_{2}^2) 
\cr
\times&&\prod_{i=1}^N\left(
(N d)^\nu \tau^{\nu+1} (x_i y_i)^{2\nu+1} e^{-N\tau (c_1 x_i^2 + c_2 y_i^2)}
 \sum_{n_i=0}^{\infty} \frac{n_i!(1-\tau)^{n_i}}{(n_i+\nu)!} 
 L_{n_i}^{\nu} (N \tau c_1 x_i^2) L_{n_i}^{\nu} (N \tau c_2 y_i^2) 
\right)\!,\nn\\
\label{E2}
\eea
with $\tau = 1-d^2/(c_1 c_2)$.
Next we include the mass $m_2$ 
into the Vandermonde determinant $\Delta_N(\{y^2\})$, and then replace
it by a determinant of Laguerre polynomials normalised to be monic
\be
\Delta_N(\{y^2\})\prod_{i=1}^N(y_i^2+m_{2}^2)
=\Delta_{N+1}((im_2)^2,\{y^2\}) 
=\det_{j,k=0,\ldots,N}\left[ (-)^jj!(N\tau c_2)^{-j}L_j^\nu(N\tau c_2y_k^2)
\right]\ ,
\label{Ldef}
\ee
where $y_0\equiv im_2$.
We observe that the Laguerre polynomials $L_j^\nu(N\tau c_2y_k)$
now all appear with their corresponding
weight function  $y_i^{2\nu+1} e^{-N\tau c_2 y_i^2}$, except for $y_0$ of 
course\footnote{Note the 
additional $\tau$ in the exponent comparing eq. (\ref{E1}) and (\ref{E2}),
coming from the identity for the $I$-Bessel function.}. 
Writing the determinant eq. (\ref{Ldef}) as a sum
over permutations we can integrate out all variables 
$y_1$ to $y_N$ successively, each 
integral killing one infinite sum over $n_i$. Thus, under the permutation each 
$L_{n_i}^{(\nu)} (N \tau c_2 y_i^2)$ gets replaced by 
$L_{n_i}^{(\nu)} (N \tau c_1 x_i^2)$ times the norm from the integration and
the remaining factor from inside the sum. We can therefore
rewrite the result again
as a determinant, with the first row containing the mass $y_0=im_2$ unchanged: 
\bea
&&E_{0,0}(s,0) \ =\ \frac{N!}{{\cal Z}_\nu^{(1+1)}}(N d)^{N\nu} \tau^{N(\nu+1)}
\int_s^\infty dx_1\ldots dx_N 
\Delta_N(\{x^2\})
\prod_{i=1}^N (x_i^2+m_{1}^2)\ x_i^{2\nu+1}  e^{-N\tau c_1 x_i^2}
\cr
&&\times
\left|
\begin{array}{ccccc}
L_0^\nu(N\tau c_2(im_2)^2)& \cdots & 
\frac{(-)^j j!}{(N\tau c_2)^{j}}L_j^\nu(N\tau c_2(im_2)^2)& \cdots& 
\frac{(-)^N N!}{(N\tau c_2)^{N}}L_N^\nu(N\tau c_2(im_2)^2)
\\
\frac{1}{2(N\tau c_2)^{\nu+1}} L_0^\nu(N\tau c_1x_1^2)& \cdots & 
\frac{(-)^jj!(1-\tau)^j}{2(N\tau c_2)^{j+\nu+1}}
L_{j}^{(\nu)} (N \tau c_1 x_1^2) & \cdots & 
\frac{(-)^N N!(1-\tau)^N}{2(N\tau c_2)^{N+\nu+1}}
L_{N}^{(\nu)} (N \tau c_1 x_1^2)
\\
\cdots & \cdots & \cdots & \cdots &\cdots\\
\frac{1}{2(N\tau c_2)^{\nu+1}} L_0^\nu(N\tau c_1x_N^2)
& \cdots & 
\frac{(-)^jj!(1-\tau)^j}{2(N\tau c_2)^{j+\nu+1}}
L_{j}^{(\nu)} (N \tau c_1 x_N^2) & \cdots & 
\frac{(-)^N N!(1-\tau)^N}{2(N\tau c_2)^{N+\nu+1}}
L_{N}^{(\nu)} (N \tau c_1 x_N^2)
\\
\end{array}
\!\right|\nn\\
&& \nn\\
=&& 
\frac{N!(N d)^{N\nu} \tau^{N(\nu+1)}
\prod_{j=0}^{N}(1-\tau)^j(N\tau c_2)^{-j}
}{{{\cal Z}_\nu^{(1+1)}}2^N(N\tau c_2)^{N(\nu+1)}}
\int_s^\infty \!\!dx_1\ldots dx_N 
\Delta_N(\{x^2\})
\prod_{i=1}^N (x_i^2+m_{1}^2)\ x_i^{2\nu+1}  e^{-N\tau c_1 x_i^2}
\nn\\
&&\times
\left|
\begin{array}{ccccc}
\Lh_0(N\tau c_2(im_2)^2)& \cdots & 
(1-\tau)^{-j} \Lh_j(N\tau c_2(im_2)^2)& \cdots& 
(1-\tau)^{-N}\Lh_N(N\tau c_2(im_2)^2)
\\
\Lh_0(N\tau c_1x_1^2)& \cdots & 
\Lh_{j} (N \tau c_1 x_1^2) & \cdots & 
\Lh_{N} (N \tau c_1 x_1^2)
\\
\cdots & \cdots & \cdots & \cdots &\cdots\\
\Lh_0(N\tau c_1x_N^2)
& \cdots & 
\Lh_{j} (N \tau c_1 x_N^2) & \cdots & 
\Lh_{N} (N \tau c_1 x_N^2)
\\
\end{array}
\right|,
\label{Edelta}
\eea
taking out common factors. Here we have defined the following notation
for monic Laguerre polynomials
\be
\Lh_n(x)\equiv (-1)^n n!\ L_n^\nu(x)\ =\ \sum_{j=0}^n (-1)^{n+j}
\frac{n!(n+\nu)!}{(n-j)!(\nu+j)!j!} \ x^j\ \ .
\label{Lmonic}
\ee
The last determinant in eq. (\ref{Edelta})
can be almost mapped to a Vandermonde determinant, using
the following identity:
\bea
&&\left|
\begin{array}{ccccc}
\Lh_0(M_2^2)& \cdots & 
\frac{1}{(1-\tau)^{j}} \Lh_j(M_2^2)& \cdots& 
\frac{1}{(1-\tau)^{N}}\Lh_N (M_2^2)
\\
\Lh_0 (X_1^2)& \cdots & 
\Lh_{j} (X_1^2) & \cdots & 
\Lh_{N} (X_1^2)
\\
\cdots & \cdots & \cdots & \cdots &\cdots\\
\Lh_0(X_N^2)
& \cdots & 
\Lh_{j} (X_N^2) & \cdots & 
\Lh_{N} (X_N^2)
\\
\end{array}
\right|=\nn\\
&&\nn\\
=&&\left| 
\begin{array}{ccccc}
\Lh_0(\frac{M_2^2}{\tau})& \cdots & 
\frac{\tau^j}{(1-\tau)^{j}} \Lh_j(\frac{M_2^2}{\tau})& \cdots& 
\frac{\tau^N}{(1-\tau)^{N}} \Lh_N(\frac{M_2^2}{\tau})\,
\\
1& \cdots & 
X_1^{2j}& \cdots & 
X_1^{2N}
\\
\cdots & \cdots & \cdots & \cdots &\cdots\\
1
& \cdots & 
X_N^{2j}& \cdots & 
X_N^{2N}
\\
\end{array}
\right|,
\label{Ldelta}
\eea
where we have defined 
\be
M_2^2\equiv N \tau c_2(im_2)^2\ \ \mbox{and} \ \ 
X_k^2\equiv N \tau c_1 x_k^2\ . 
\label{M2def}
\ee
A proof of this relation is given in 
appendix \ref{detid}.

We can now change variables $x_j\to u_j=x_j^2$, and perform the shift 
$u_j\to z_j=u_j-s^2$ to obtain integrations $\int_0^\infty dz_j$ in
eq. (\ref{Edelta}). The change of variables and subsequent
shifts induce the following changes: 
\bea
\Delta_N(\{u\})   &\to&  \Delta_N(\{z\})\ ,\nn\\
(u_j+m_1^2)       &\to& (z_j+s^2+ m_1^2) \equiv (z_j + m_1^{\prime\,2})\ ,\nn\\
u_j^\nu           &\to& (z_j+s^2)^\nu \ ,\nn\\
e^{-N\tau c_1 u_i}&\to& e^{-N\tau c_1 z_i} e^{-N\tau c_1 s^2}\ .
\label{shift}
\eea
In other words: the Vandermonde determinant
remains invariant, the mass $m_1$ receives a shift to 
$m_1^{\prime\,2}=s^2+ m_1^2$, the topology term becomes a $\nu$-fold
degenerate mass term with mass $s^2$, and the weight is shifted by a constant
factor. While this is just as in the chiral one-matrix theory, the difference
here is that the almost Vandermonde eq. (\ref{Ldelta}) is not invariant, and
becomes 
\bea
&&\left| 
\begin{array}{ccccc}
\Lh_0(\frac{M_2^2}{\tau})& \cdots & 
\frac{\tau^j}{(1-\tau)^{j}} \Lh_j (\frac{M_2^2}{\tau})& \cdots& 
\frac{\tau^N}{(1-\tau)^{N}} \Lh_N (\frac{M_2^2}{\tau})
\\
1& \cdots & 
X_1^{2j}& \cdots & 
X_1^{2N}
\\
\cdots & \cdots & \cdots & \cdots &\cdots\\
1
& \cdots & 
X_N^{2j}& \cdots & 
X_N^{2N}
\\
\end{array}
\right|=
\label{Ldeltashift}
\\
=&&
\left| 
\begin{array}{ccccc}
\Lh_0(\frac{M_2^2}{\tau})& \cdots & 
\frac{\tau^j}{(1-\tau)^{j}} \Lh_j(\frac{M_2^2}{\tau})& \cdots& 
\frac{\tau^N}{(1-\tau)^{N}} \Lh_N(\frac{M_2^2}{\tau})
\\
1& \cdots & 
(N \tau c_1(z_1+s^2))^{j}& \cdots & 
(N \tau c_1(z_1+s^2))^{N}
\\
\cdots & \cdots & \cdots & \cdots &\cdots\\
1
& \cdots & 
(N \tau c_1(z_N+s^2))^{j}& \cdots & 
(N \tau c_1(z_N+s^2))^{N}
\\
\end{array}
\right|\nn\\
=&& 
\left|
\begin{array}{ccccc}
\Lh_0(\frac{M_2^2}{\tau})
&\cdots & 
\sum_{l=0}^j\frac{\tau^l}{(1-\tau)^{l}} \Lh_l(\frac{M_2^2}{\tau})(-S^2)^{j-l}
{j \choose l}
& \cdots & 
\sum_{l=0}^N\frac{\tau^l}{(1-\tau)^{l}} \Lh_l (\frac{M_2^2}{\tau})(-S^2)^{N-l}
{N \choose l}
\\
1& \cdots & 
Z_1^{j}& \cdots & 
Z_1^{N}
\\
\cdots & \cdots & \cdots & \cdots &\cdots\\
1
& \cdots & 
Z_N^{j}& \cdots & 
Z_N^{N}
\\
\end{array}
\right|,\nn
\eea
where we have defined 
\be
S^2\equiv N \tau c_1 s^2\ \ \mbox{and} \ \ 
Z_k\equiv N \tau c_1 z_k\ . 
\ee
Here 
we have subsequently added columns\footnote{Usually one goes from monic 
powers to monic polynomials in
  this step. It is easy to invert this by defining $Z_i^\prime\equiv Z_i+S^2$ 
and then going from monic powers $Z_i^{\prime\, j}$ to  
polynomials $(Z_i^\prime-S^2)^j=Z_i^j$.}  to turn back to monic
powers in $Z_k$. The fact that it is not quite invariant illustrates
the fundamental property that there is apparently no way to map the present
two-matrix problem onto an equivalent one-matrix problem.

In the next step we turn the monic powers $Z_i^j$
back into monic Laguerre polynomials,
using again the invariance of the determinant. This will introduce yet another
sum over the Laguerre polynomials in the mass variable $M_2^2$ in the first
row. 
Because of the shift eq. (\ref{shift}) we only need to generate Laguerre
polynomials with topology $\nu=0$,  $\hat{L}_j^{\nu=0}\equiv \hat{L}_j$,
to obtain the polynomials orthogonal 
to the shifted weight. We thus obtain for the last determinant in
eq. (\ref{Ldeltashift})  
\be
\left| 
\begin{array}{ccccc}
Q_0(M_2^2) 
&\cdots &
Q_j(M_2^2)
& \cdots &Q_N(M_2^2)
\\
\hat{L}_0(Z_1) & \cdots & 
\hat{L}_j(Z_1)& \cdots & \hat{L}_N(Z_1)
\\
\cdots & \cdots & \cdots &\cdots &\cdots\\
\hat{L}_0(Z_N) & \cdots & 
\hat{L}_j(Z_N)& \cdots & \hat{L}_N(Z_N)
\\
\end{array}
\right|,
\label{Lfinal}
\ee
which defines {\em new polynomials}
\bea
Q_n(M_2^2)&\equiv&
\sum_{j=0}^n \frac{(-)^{n+j}(n!)^2}{(n-j)!(j!)^2} 
\sum_{l=0}^j\frac{\tau^l}{(1-\tau)^{l}} 
\hat{L}_l\left(\frac{M^2_2}{\tau}\right)(-S^2)^{j-l}
{j \choose l}\nn\\
&=& (-)^n n!
\sum_{l=0}^n \frac{\tau^l}{(1-\tau)^{l}} {L}_l\left(\frac{M^2_2}{\tau}\right)
L_{n-l}^l(-S^2)\nn\\
&=& (-)^n n!\sum_{l=0}^n L_{n-l}^l(-S^2)\sum_{k=0}^l
\frac{(-)^{k+l}l!}{(l-k)!k!(1-\tau)^k}L_k(M_2^2)\nn\\
&=& (-)^n n!
\sum_{k=0}^n
\frac{1}{(1-\tau)^k}L_k(M_2^2) L_{n-k}^{-1}(-S^2)
\ .
\label{Q}
\eea
In the first step we have swapped sums, $\sum_{j=0}^n\sum_{l=0}^j\to
\sum_{l=0}^n\sum_{j=l}^n$, such that the powers in $S^2$ give the 
Laguerre polynomial $L_{n-l}^l(-S^2)$. In the second step we have used the
identity (\ref{Lid2}) backwards in order to take the argument $1/\tau$ out of
the first Laguerre polynomial, in choosing $z=M_2^2$ and $w=1/\tau$ in eq. 
(\ref{Lid2}). This goes at the expense of introducing another sum. 
After swapping again sums to $\sum_{k=0}^n\sum_{l=k}^n$, the latter sum over
the generalised Laguerre polynomial in $-S^2$ can be simplified, using the
following identity, 
\be
L_{n-k}^{-1}(-S^2)\ =\ 
\sum_{j=0}^{n-k}(-)^j{j+k \choose k} L_{n-k-j}^{j+k}(-S^2) \ . 
\label{LidII}
\ee
A proof of this simple identity is presented in the appendix \ref{B}. 
Note that all polynomials $L_{n-k}^{-1}(-S^2)$ are
proportional to $-S^2$, except for $n=k$ as $L_0^{-1}(-S^2)=1$. 
This will become important when computing the normalisation in the limit
$s\to0$.

The explicit appearance of a new set of polynomials is again a reminder
that we cannot map the problem onto a one-matrix problem. Of course,
in the limit of the deformation parameters $\mu_{1,2} \to 0$, the polynomials
$Q_n(M_2^2)$ reduce to Laguerre polynomials.
In the form given in the last line of eq. (\ref{Q}) the new polynomials
$Q_n$ are amenable to the large-$N$ scaling limit that we take in the next
subsection.

Returning to the gap probability, 
in the last step we now replace in eq. (\ref{Edelta}) the determinant
$\Delta_N(\{x^2\})=\Delta_N(\{z\})$
times the mass term 
by a larger Vandermonde determinant, 
and then replace monic powers by Laguerre polynomials
monic in the arguments $z_k$. 
The $\nu$ degenerate masses obtained after the shift
eq. (\ref{shift})
can be dealt with by first taking
them different, and then taking limits by multiple application of 
l'H\^opital's rule. For simplicity we set $\nu=0$ in
all of the following. We have  
\be
\Delta_N(\{z\})\prod_{i=1}^N (z_i + m_1^{\prime\,2})
=\Delta_{N+1}((im_1^\prime)^2,\{z\}) 
=\det_{j,k=0,\ldots,N}\left[ (-)^jj!(N\tau c_1)^{-j}L_j(N\tau c_1z_k)
\right]\ ,
\label{Ldef2}
\ee
where we define $z_0=im_1^\prime$. 

Let us collect what we have derived so far:
\bea
&&E_{0,0}(s,0)\ =\ \frac{N!\prod_{j=0}^{N}(1-\tau)^j(N\tau c_2)^{-j}
(N\tau c_1)^{-j}}{{{\cal Z}_0^{(1+1)}}\ 2^{2N}(Nc_2)^{N}}
e^{-N^2\tau c_1s^2}
\int_0^\infty dz_1\ldots dz_N\ e^{-N\sum_{i=0}^N \tau c_1 z_i}
\nn\\
&&\times
\left| 
\begin{array}{ccccc}
\hat{L}_0(M_1^{\prime\,2}) 
&\cdots &
\hat{L}_j(M_1^{\prime\,2})
& \cdots &\hat{L}_N(M_1^{\prime\,2})
\\
\hat{L}_0(Z_1) & \cdots & 
\hat{L}_j(Z_1)& \cdots & \hat{L}_N(Z_1)
\\
\cdots & \cdots & \cdots &\cdots &\cdots\\
\hat{L}_0(Z_N) & \cdots & 
\hat{L}_j(Z_N)& \cdots & \hat{L}_N(Z_N)
\\
\end{array}
\right|\left| 
\begin{array}{ccccc}
Q_0(M_2^2) 
&\cdots &
Q_j(M_2^2)
& \cdots &Q_N(M_2^2)
\\
\hat{L}_0(Z_1) & \cdots & 
\hat{L}_j(Z_1)& \cdots & \hat{L}_N(Z_1)
\\
\cdots & \cdots & \cdots &\cdots &\cdots\\
\hat{L}_0(Z_N) & \cdots & 
\hat{L}_j(Z_N)& \cdots & \hat{L}_N(Z_N)
\\
\end{array}
\right|.
\nn\\
\label{Efull}
\eea
Here the definition 
\be
M_1^{\prime\,2}\equiv N\tau c_1(im_1^\prime)^2=
-N\tau c_1(m_1^2+s^2)
\label{M1primedef}
\ee
has been used.
We can now apply the orthogonality of the Laguerre polynomials with respect to
the weight $e^{-N\tau c_1 z_i}$ to compute the $N$-fold integral over the
determinants, applying the standard Dyson Theorem.
We thus obtain the final answer for finite $N$:
\bea
E_{0,0}(s,0) &=& C\,
e^{-N^2\tau c_1s^2}{K}_{N+1} (M_1^{\prime\,2},M_2^2)\ ,
\nn\\
{K}_{N+1} (M_1^{\prime\,2},M_2^2) &\equiv&
\sum_{j=0}^N \frac{(-)^j}{j!} L_j(M_1^{\prime\,2})
Q_j(M_2^2)\ .
\label{Efinal}
\eea
This result defines a {\it new  
kernel} of the polynomials $L_j$ and $Q_j$ in the
(shifted) masses.  The constant 
$C\equiv 1/\sum_{j=0}^N \frac{1}{(1-\tau)^j} L_j(M_1^{2})
L_j(M_2^2)$ that is inversely proportional to
the partition function ${\cal Z}^{(1+1)}_0$
ensures the correct normalisation $\lim_{s\to0}E_{0,0}(s,0)=1$.
It can be obtained independently by computing the partition function
\be
{{\cal Z}_0^{(1+1)}}\ =\ 
\frac{N!^2\prod_{j=0}^{N}(j!)^2(1-\tau)^j(N\tau c_1)^{-j}(N\tau c_2)^{-j}
}{2^{2N}(Nc_1)^{N}(N\tau c_2)^{N}}
\sum_{j=0}^N \frac{1}{(1-\tau)^j} L_j(M_1^{2})
L_j(M_2^2)
\ ,
\ee
following the
same steps as before but setting $s=0$. The calculation simplifies in eq. 
(\ref{Ldeltashift}) so that the $Q_j$'s become Laguerre polynomials.
Indeed as a check we can take 
\be
\lim_{s\to0}Q_n(M_2^2)\ =\ \frac{(-)^n n!}{(1-\tau)^n} L_n(M_2^2)\ ,
\label{lims0Qn}
\ee
where only the term $L_{n-k=0}^{-1}(-S^2)=1$ contributes to the sum. 
This already indicates that this last term in the sum is special.

As a further independent check we may take the limit $\mu_1,\mu_2\to0$
($\Rightarrow\tau\to0$). In 
this limit the two Dirac operators become equal, ${\cal D}_1={\cal D}_2$, 
and we 
should recover the known one-matrix theory result \cite{DNW}. Indeed, we get
\bea
\lim_{\mu_1,\,\mu_2\to0}\frac{(-)^n}{n!}Q_n(M_2^2) 
&=&
\sum_{k=0}^nL_k(-Nm_2^2) L_{n-k}^{-1}(-Ns^2)
\label{Lsumid}\\
&=& 
\sum_{j=0}^n \frac{(-)^{j}n!}{(n-j)!(j!)^2} 
\sum_{l=0}^jM^{2l}_2(-S^2)^{j-l}
{j \choose l}
\ =\ L_n\Big(-N(m_2^2+s^2)\Big)\ ,
\nn
\eea
$i.e.$, Laguerre polynomials of shifted mass just as for the first flavour
$m_1^\prime$. This follows from the first and last line of the 
definition eq. (\ref{Q}).
Inserted into the kernel eq. (\ref{Efinal}) we obtain the
one-matrix theory result for the gap probability in terms of 
the partition function of 2 flavours with shifted masses.

It is straightforward to see that for more flavours, 
$N_1>1$ and $N_2>1$, the very same steps still
go through (see also the corresponding
determinant identity in appendix \ref{detid}). 
The only difference is that the there will be more rows with masses of flavour
$N_1$ in the determinant eq. (\ref{Ldef2}), and more masses of flavour 
$N_2$ in the new
polynomials $Q_j$ in eq. (\ref{Lfinal}). The absorption of the mass terms into
a larger Vandermonde determinant
leads to inverse Vandermonde determinants in each of the $N_1$ and $N_2$ 
masses, which can be taken out of the integral. We arrive at 
\bea
&&E_{0,0}(s,0)\ \sim\ \frac{1}{
{\cal Z}_0^{(N_f)}\Delta_{N_1}(m_{f1}^2)\Delta_{N_2}(m_{f2}^2)}
e^{-N^2\tau c_1s^2}
\int_0^\infty dz_1\ldots dz_N\ e^{-N\sum_{i=0}^N \tau c_1 z_i}
\label{EfullNf}\\
&&\times
\left| 
\begin{array}{ccc}
\hat{L}_0(M_{f1=1}^{\prime\,2}) 
& \cdots &\hat{L}_{N+N_1-1}(M_{f1=1}^{\prime\,2})
\\
\cdots & \cdots  &\cdots\\
\hat{L}_0(M_{N_1}^{\prime\,2}) 
& \cdots &\hat{L}_{N+N_1-1}(M_{N_1}^{\prime\,2})
\\
\hat{L}_0(Z_1) & \cdots & 
\hat{L}_{N+N_1-1}(Z_1)
\\
\cdots & \cdots  &\cdots\\
\hat{L}_0(Z_N) & \cdots & 
\hat{L}_{N+N_1-1}(Z_N)
\\
\end{array}
\right|\left| 
\begin{array}{ccccc}
Q_0(M_{f2=1}^2) 
& \cdots &Q_{N+N_2-1}(M_{f2=1}^2)
\\
\cdots & \cdots&\cdots\\
Q_0(M_{N_2}^2) 
& \cdots &Q_{N+N_2-1}(M_{N_2}^2)
\\
\hat{L}_0(Z_1) & \cdots & 
\hat{L}_{N+N_2-1}(Z_1)
\\
\cdots & \cdots &\cdots\\
\hat{L}_0(Z_N) & \cdots & 
\hat{L}_{N+N_2-1}(Z_N)
\\
\end{array}
\right|.
\nn
\eea 
The orthogonality of Laguerre polynomials can be exploited in the manner
of ref. \cite{AV}. This leads to the following determinant expressions. 
For an equal number of flavours $N_1=N_2$ we have\footnote{We omit all
mass dependent normalisation constants here that can be obtained easily. In
particular they will cancel the Vandermonde determinants of the masses, see
e.g. \cite{ADOS} for the partition functions.} 
\be
E_{0,0}(s,0)= const. 
\frac{1}{{\cal Z}_0^{(N_f)}
\Delta_{N_1}(\{m_{f1}^2\})\Delta_{N_2}(\{m_{f2}^{2}\})}\ 
e^{-N^2\tau c_1s^2}\det_{1\leq f1,f2\leq N_1}
\left[ {K}_{N+N_1-1} (M_{f1}^{\prime\,2},M_{f2}^2)\right]\ . 
\label{EN1=N2}
\ee
In the case where $N_1(N_2)$ is larger, the determinant is of size $N_1(N_2)$
and contains additional polynomials $Lj(M_{j1}^{\prime\,2})\
(Q_j(M_{k2}^2))$ to fill up the additional columns (rows) \cite{AV},
\newpage
\bea
N_1>N_2:&&E_{0,0}(s,0)= const. 
\frac{1}{{\cal Z}_0^{(N_f)}
\Delta_{N_1}(\{m_{f1}^2\})\Delta_{N_2}(\{m_{f2}^{2}\})}\ 
e^{-N^2\tau c_1s^2}\label{EN1>N2}\\
\times&&\det_{f1,f2}\left[ 
{K}_{N+N_2-1} (M_{f1}^{\prime\,2},M_{f2=1}^2) \cdots 
{K}_{N+N_2-1} (M_{f1}^{\prime\,2},M_{N_2}^2) 
\,{L}_{N+N_2}(M_{f1}^{\prime\,2})\cdots 
{L}_{N+N_1-1}(M_{f1}^{\prime\,2})
\right]\nn
\eea
where we display the $f1$-th row in mass $M_{f1}^{\prime\,2}$ of flavour
$N_1$, and 
\bea
N_2>N_1:&&E_{0,0}(s,0)= const. 
\frac{1}{{\cal Z}_0^{(N_f)}
\Delta_{N_1}(\{m_{f1}^2\})\Delta_{N_2}(\{m_{f2}^{2}\})}\ 
e^{-N^2\tau c_1s^2}\label{EN1<N2}
\\
\times&&\det_{f2,f1}\left[ 
{K}_{N+N_1-1} (M_{f2}^{2},M_{f1=1}^{\prime\,2}) \cdots 
{K}_{N+N_1-1} (M_{f2}^{2},M_{N_1}^{\prime\,2}) 
\,Q_{N+N_1}(M_{f2}^{2})\cdots 
Q_{N+N_2-1}(M_{f2}^{2})
\right]\nn
\eea
Here we have transposed the matrix to 
display the $f2$-th row in mass $M_{f2}^{2}$ of flavour $N_2$.  

For example,
this includes in particular the interesting case of quenching the
first flavour ($i.e.$ putting $N_1=0$) while keeping its  chemical potential
nonzero, $\mu_1\neq 0$. This quenched
flavour can then be measured in the background
of $N_2=2$ flavours with masses $m_1$ and $m_2$ ,
\be
N_2=2,N_1=0:\ \ E_{0,0}(s,0)= const. 
\frac{e^{-N^2\tau c_1s^2}}{{\cal Z}_0^{(2)}(m_2^2-m_1^2)} 
\det\left[
\begin{array}{cc} 
Q_{N}(-N\tau c_2m_1^{2})& Q_{N+1}(-N\tau c_2m_1^{2})\\
Q_{N}(-N\tau c_2m_2^{2})& Q_{N+1}(-N\tau c_2m_2^{2})\\
\end{array}
\right].
\label{N2=2pq}
\ee
In particular, setting the chemical potential 
of the dynamical flavours $N_2$ to zero, $\mu_2=0$,
will not eliminate the
other chemical potential $\mu_1\neq0$ (see eq. (\ref{cdtdef})), 
or reduce to a known one-matrix quantity. This non-trivial
$\mu$-dependence due to the valence quarks can serve as
a clean way to measure the pion decay constant $F_{\pi}$
from gauge field ensembles generated with dynamical light
quarks that carry no chemical potential.

Finally, as was pointed out earlier, the probability
corresponding to non-vanishing
gauge field topology $\nu\neq0$ can be
introduced by adding $\nu$ extra masses of ${\cal D}_1$, and then taking them
to be degenerate with value $s^2$.

\subsection{The large-$N$ limit}

In this subsection we take the large-$N$ scaling
limit by the same rescaling as 
in ref. \cite{ADOS}, to which we refer for more details. We first derive
the limits of all building blocks needed for the general case, and then
specify the fully explicit result in three examples in subsequent subsections.

All eigenvalues, the gap and the masses are rescaled in the same way 
(as would be $\hy=2Ny$), 
the usual microscopic limit
\bea
\hx &\equiv& 2Nx\ ,\ \ \hs\ \equiv\ 2Ns\ ,\nn\\
\hm_f&\equiv& 2Nm_f \ ,\ \ \hmu_f \ \equiv\ 2N\mu_f^2\ \ \mbox{for} \ \
f=1,2\ ,\nn\\ 
\hd &\equiv& \hmu_2-\hmu_1 \ .
\label{scaledef}
\eea
All scalings including the chemical potential keeping $N\mu_f^2$ fixed 
can be read off from the chiral Lagrangian
eq. (\ref{ZXPT}).
For the various constants containing the $\mu_f$ this implies the following
scaling:
\bea
&&\lim_{N,j,k\to\infty} (1-\tau)^{-k}= \exp\left[\frac12 rt\hd^2 \right] 
\ \ \mbox{where}\ \  
t\equiv j/N, \ \ r\equiv k/j\ , \nn\\
&& \lim_{N\to\infty} \tau c_f=1 \ \ \mbox{for}\ \ f=1,2\ .
\label{scale2}
\eea
For the Laguerre polynomials the following scaling holds:
\bea
&&\lim_{N,j\to\infty}L_{j}(M_1^{'\,2}=-N\tau c_1m_1^{\prime\,2})
\ =\ I_0(\sqrt{t}\ \hm_1')\ ,
\label{Lasymp}\\
&&\lim_{N,j,k\to\infty} L_{j-k\neq0}^{-1}(-S^2=-N\tau c_1s^2)\ =\  
\frac{1}{2j}\sqrt{\frac{t}{(1-r)}}\ \hs\ I_{1}(\sqrt{(1-r)t}\ \hs),
\nn
\eea
recalling $\hm_1^{'\,2}=\hm_1^2+\hs^2$. Special care has to be
taken in the asymptotic of the new polynomial, 
\bea
\lim_{j\to\infty}\frac{(-)^j}{j!} Q_j(M_2^2=-N\tau c_2m_2^2)&=& 
\lim_{j\to\infty}
\left(\sum_{k=0}^{j-1}\frac{1}{(1-\tau)^k}L_k(M_2^2)
\ L_{j-k}^{-1}(-S^2)\ +\ \frac{1}{(1-\tau)^j}
L_j(M_2^2)\cdot 1
\right)\nn\\
\Leftrightarrow \ \ \ \ Q_S(\hm_2;t)
&\equiv& \frac12\int_0^1dr\ e^{\frac12 rt \hd^2} I_0(\sqrt{rt}\ \hm_2) 
\sqrt{\frac{t}{1-r}}\ \hs\ I_{1}(\sqrt{(1-r)t}\ \hs)
\nn\\
&&+\ e^{\frac12 t \hd^2}
I_0(\sqrt{t}\ \hm_2) 
\ .
\label{Qasymp}
\eea
Here we have to split off the $s$-independent part $L_0^{-1}=1$, 
which is the single term surviving in the limit $s\to 0$, and
which hence ensures the
normalisability of the probability in that limit. Usually neglecting 
a single term when replacing a sum by an integral amounts to removing a
quantity of measure zero, which should be 
irrelevant. However, in our case this is not
true as the convergence in $\hs$ is not uniform. Therefore we have to treat
that term separately and find the ``anomalous'' $I_0$-term in the
scaling limit. That this procedure is correct is checked by
computing the normalisation before and after taking the large-$N$ limit.
This curious phenomenon together
with the appearance of Laguerre polynomials $L_{j-k}^{-1}$ leads to our 
new microscopic kernel.

The final answer for the microscopic limit of the new  kernel in
eq. (\ref{Efinal}) thus reads
\bea
K_S(\hm_1',\hm_2)&\equiv& 
\lim_{N\to\infty}\frac{1}{N}K_{N+1}(M_{1}^{\prime\,2},M_{2}^2)
\label{KS}\\
&=& 2\int_0^1dTT^2 I_0(T\hm_1') \int_0^1dRR\frac{1}{\sqrt{1-R^2}}\ 
e^{\frac12 R^2T^2 \hd^2} I_0(RT\hm_2)\   
\hs\ I_{1}(\hs T\ \sqrt{1-R^2}) \nn\\
&&+\ 2\int_0^1dTT\ e^{\frac12 T^2\hd^2}I_0(T\hm_1')I_0(T\hm_2)
\ ,
\nn
\eea
where we have changed to squared variables. This kernel can no longer be
related to a single partition functions of shifted masses, as it was the case
in the one-matrix theory \cite{DNW}.

Likewise we obtain for the 
normalisation constant which is proportional to the partition function,
\be
\lim_{N\to\infty}\frac{1}{N}\sum_{j=0}^N \frac{1}{(1-\tau)^j} L_j(M_1^{2})
L_j(M_2^2)\ =\ 2\int_0^1dTT \exp\left[\frac12 T^2
  \hd^2\right]I_0(T\hm_1)I_0(T\hm_2) \ .
\ee
Note that the first mass $\hm_1$ is {\it not} shifted here, in contrast to the
previous equation. 
Partition functions of more flavours follow easily given the building blocks
above, together with the general expressions given in \cite{AV,ADOS}.
We now have all ingredients to obtain 
all gap probabilities with any flavour content by inserting the
asymptotic kernel eq. (\ref{KS}) and asymptotic polynomials eqs. (\ref{Lasymp})
and (\ref{Qasymp}) into the respective eqs. (\ref{EN1=N2}) --  (\ref{EN1<N2}),
normalised by the corresponding partition function. 

In the following we give three simple examples that illustrate these 
very general expressions. In order to guide the eye we mostly display the
distribution of the first eigenvalues versus the corresponding eigenvalue
density it has to follow. For comparison to Lattice results the gap
probability that we give explicitly may be even more useful as it allows for a
binning independent comparison with data.


\subsection{Two light flavours}

\begin{figure*}[ht]
  \unitlength1.0cm
  \epsfig{file=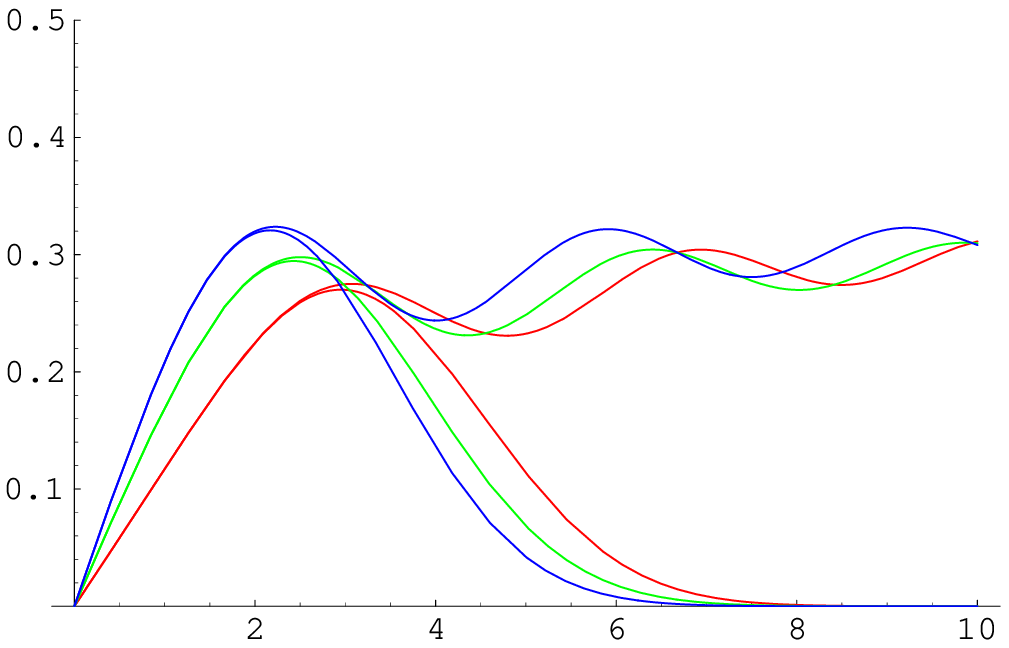,clip=,width=10cm}
  \caption{ 
    \label{fig:1} 
The eigenvalue density and first eigenvalue for $N_f=1+1$ 
with imaginary
    chemical potential $\hd=1$ (low red), 3( middle green), and 10 (upper
    blue 
    curve), at fixed quark masses
    $\hat{m}_1=3$, $\hat{m}_2=4$.
    }
\end{figure*}

Let us first consider
the gap probability corresponding to two flavours $N_1=N_2=1$, as given in
eq. (\ref{Efinal}). Collecting the formulae from above we obtain  
\bea
\lim_{N\to\infty}E_{0,0}(s,0) &\equiv& E_{S\ 0,0}^{(1+1)}(\hs,0)
\ =\ \left(\int_0^1dTT e^{\frac12 T^2 \hd^2}I_0(T\hm_1)
I_0(T\hm_2)\right)^{-1}\exp\left[-\frac14\hs^2\right]\nn\\
&\times&\left( \int_0^1dTT^2 I_0(T\hm_1') \int_0^1dRR\frac{1}{\sqrt{1-R^2}}\ 
e^{\frac12 R^2T^2  \hd^2} I_0(RT\hm_2)\   
\hs\ I_{1}(\hs T\sqrt{1-R^2}) \right.\nn\\
&&\left.
+\int_0^1dTT e^{\frac12 T^2  \hd^2}I_0(T\hm_1')I_0(T\hm_2)\right).
\label{E2MM11}
\eea
Once more we can perform an analytic check by taking $\hd\to0$ in order to
go back to the known one-matrix quantity \cite{DNW}. Using the following
so-called Sonine integral identity \cite{Prudnikov}
(that also follows from the large-$N$ limit
of identity eq. (\ref{Lsumid}))
\be
s\int_0^1dx\frac{x}{\sqrt{1-x^2}}\ I_0(mx)I_1(s\sqrt{1-x^2}) \ +\ I_0(m)
\ =\ I_0(\sqrt{m^2+s^2})\ ,
\label{Sonine}
\ee
we obtain the known gap probability \cite{DNW} 
as a ratio of a two-flavour partition
function with shifted masses over one with unshifted ones\footnote{The
  remaining integral is elementary and gives 
a $2\times2$ determinant of Bessel
  functions.}. 
As an illustration, we show the distribution of the first eigenvalues 
$p_{S\ 1,0}^{N_f=1+1}(\hs,0)=-\partial_{\hs}E_{S\ 0,0}^{N_f=1+1}(\hs,0)$
for different values of $\hd$ and compare it to the corresponding densities in
fig. \ref{fig:1}. 
From \cite{ADOS} we have for the density
\bea
\rho_{1,0}^{(1+1)}(\hx)&=& \rho_{1M\!M}^Q(\hx)-\hx \ 
\frac{\int_0^1dttJ_0(t\hx)I_0(t\hm_1)\int_0^1dtt\ e^{\frac12 t^2  \hd^2}
J_0(t\hx)I_0(t\hm_2)}{
\int_0^1dtt e^{\frac12 t^2  \hd^2}I_0(t\hm_1)I_0(t\hm_2)}\ ,
\label{rho2MMN11}
\eea
where we have introduced the one-matrix model quenched density 
$\rho_{1M\!M}^Q(\hx)$ 
from eq. (\ref{rhoQ}) below.

\begin{figure*}[ht]
  \unitlength1.0cm
  \epsfig{file=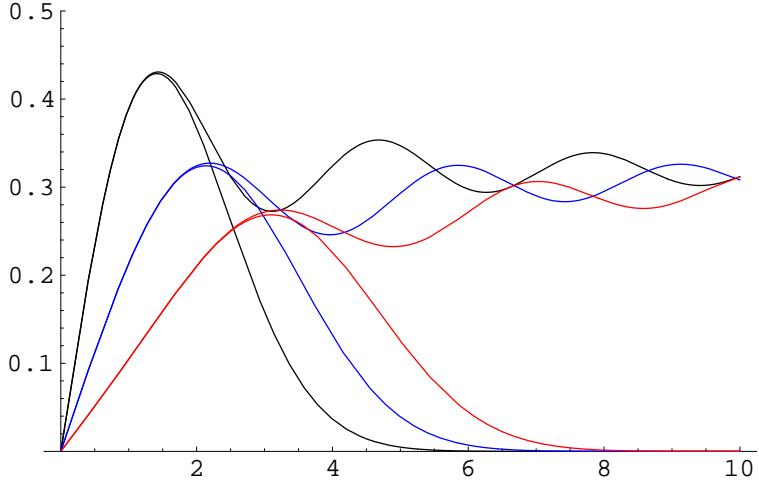,clip=,width=10cm}
  \caption{ 
    \label{fig:2} 
The eigenvalue density and first eigenvalue of the one-matrix
theory: two flavours with $\hat{m}_1=3$, $\hat{m}_2=4$ 
(low red), one flavour with $\hat{m}_1=3$ (middle blue),
and the quenched case (upper
    black    curve).
}
\end{figure*}

At $\hd=0$ eq. (\ref{rho2MMN11}) coincides with the corresponding one-matrix
model density eq. (\ref{rho1MMN2}) below. 
For $\hd=1$ the curve is still close to this density, compare to 
fig. \ref{fig:2}. 
For
$\hd\gg1$ the curves approach the one-matrix quantities of 
{\it one flavour $N_1=1$} with mass  $\hat{m}_1=3$ (the flavour 
corresponding to the $y$-eigenvalues gets
quenched), compare again to fig. \ref{fig:2}
below.
This fact can be seen analytically, by taking the limit $\hd\to\infty$
in eq. (\ref{rho2MMN11}) and doing a saddle point approximation,
\be
\lim_{\hd\to\infty}
\int_0^1dtt\ e^{\frac12 t^2  \hd^2} J_0(t\hx)I_0(t\hm_2)\ \sim\ 
\hd^{-2}e^{\frac12 \hd^2} J_0(\hx)I_0(\hm_2)\ .
\label{SPint}
\ee
The $\hd$-dependent integrals get replaced by their
values at the upper limit $t=1$, and we have also computed the subleading
coefficient for later convenience.
After cancelling common factors in eq. (\ref{rho2MMN11}) we obtain,
as we should, the 
one-matrix density for one flavour with mass  $\hat{m}_1$ as in
eq. (\ref{rho1MM1}) below.
We have checked that the same limit applies to the first eigenvalue
distribution.

For the comparison above we give the following known one-matrix quantities 
\cite{DNW,DN} that are 
displayed in figure \ref{fig:2}.
The quenched density and its first eigenvalues read
\be
\rho^Q_{1M\!M}(\hx) \ =\ \frac{\hx}{2}\left(J_0(\hx)^2+J_1(\hx)^2\right)\ ,
\ \ \ \ p^Q_{1M\!M}(\hs) \ =\ \frac12\hs\ e^{-\frac14 \hs^2} \ .
\label{rhoQ}
\ee
The massive two-flavour density is given by 
\bea
\rho_{1M\!M}^{(2)}(\hx)&=& \rho^Q_{1M\!M}(\hx)-\hx \ 
\frac{\int_0^1dttJ_0(t\hx)I_0(t\hm_1)\int_0^1dtt
J_0(t\hx)I_0(t\hm_2)}{
\int_0^1dtt I_0(t\hm_1)I_0(t\hm_2)}
\label{rho1MMN2}
\eea
as well as its first eigenvalue distribution by 
\bea
p_{1M\!M}^{(2)}(\hs)&=& \frac12\hs\ e^{-\frac14 \hs^2}\ 
\frac{I_2(\hm_1') \hm_2' 
I_3(\hm_2') -I_2(\hm_2')  
\hm_1' 
I_3(\hm_1')}{
        I_0(\hm_1)\hm_2I_1(\hm_2)-I_0(\hm_2)\hm_1 I_1(\hm_1)}\ . 
\label{p1MMN2}
\eea
Here primed masses are shifted according to
$\hm_i^{'\,2} \equiv \hm_i^2 + s^2$. 
We also need the one-flavour 
density and its first eigenvalue
\bea
\rho_{1M\!M}^{(1)}(\hx) &=& \rho^Q_{1M\!M}(\hx)\ -\ 
\hx J_0(\hx)\ \frac{\int_0^1dttJ_0(t\hx)I_0(t\hm_1)}{I_0(\hm_1)}\ ,
\label{rho1MM1}\\
p_{1M\!M}^{(1)}(\hs) &=& \frac12\hs\ e^{-\frac14 \hs^2}
\frac{I_2(\hm_1^{'\,2})}{I_0(\hm_1)}\ .
\label{p1MM1}
\eea


\subsection{Partial quenching}

\begin{figure*}[ht]
  \unitlength1.0cm
  \epsfig{file=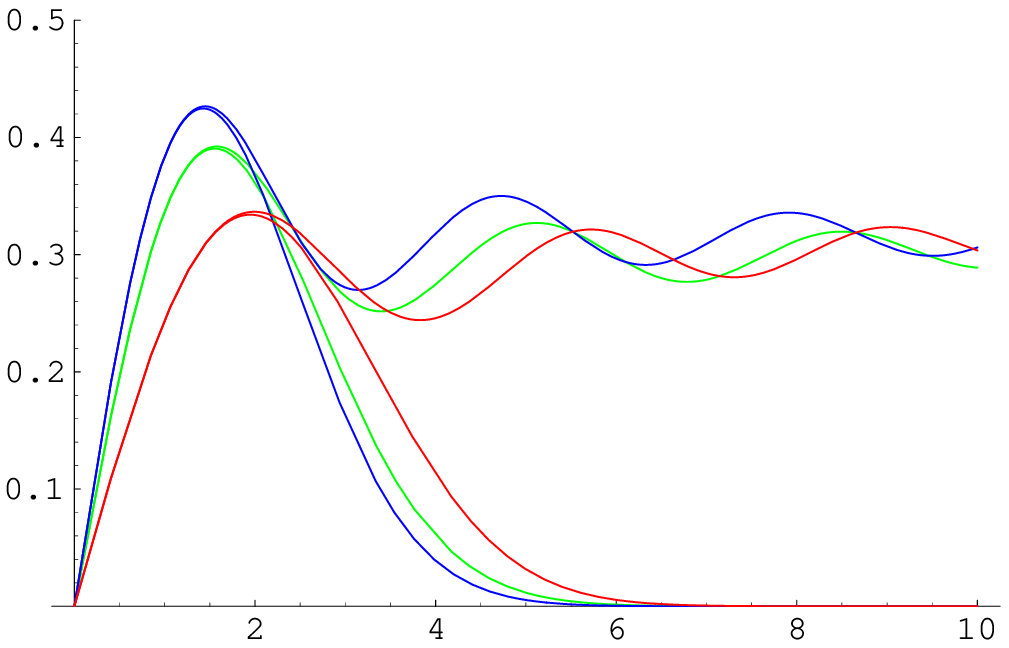,clip=,width=10cm}
  \caption{ 
    \label{fig:3} The eigenvalue density and first eigenvalue for $N_f=0+1$ 
with imaginary
    chemical potential $\hd=1$ (low red), 3( middle green), and 10 (upper
    blue 
    curve) and fixed quark mass
    $\hat{m}_1=3$.
}
\end{figure*}
As another example we can consider the 
partially quenched gap probability eq. (\ref{N2=2pq})
with $N_1=0$ and one single ($N_2=1$) flavour of mass $m_1$,
\be
E_{S\ 0,0}^{(0+1)}(\hs,0)\ \equiv\ \exp\left[-\frac14\hs^2- \frac12\hd^2\right]
\frac{Q_S(\hm_1;t=1)}{I_0(\hm_1)} \ ,
\label{E2MM01}
\ee
where 
the extra $\hd$-dependent factor $e^{-\hd^2/2}$ comes from the
partition function ${\cal Z}_0^{(0+1)}$ that normalises this gap
probability.
Its derivative 
is shown in fig. \ref{fig:3} together with the corresponding density
\cite{ADOS}
\be
\rho_{1,0}^{(0+1)}(\hx)\ =\ \rho^Q_{1M\!M}(\hx)\ -\ 
\exp\left[-\frac12\hd^2\right]\hx
\frac{ J_0(\hx)}{I_0(\hm_1)} \int_0^1dTT e^{\frac12 T^2 \hd^2}I_0(T\hm_1)
J_0(T\hx)\ .
\label{rho2MM01}
\ee
Again we recover the one-matrix density eq. (\ref{rho1MM1}) when setting
$\hd=0$ in eq. (\ref{rho2MM01}).  
For $\hd=1$ the curve is still close to this one-flavour
one-matrix result, see fig. \ref{fig:2}, 
and for
$\hd\gg1$ the curves approach the quenched one-matrix density. 
This can again be checked analytically by taking $\hd\to\infty$ of eq. 
(\ref{rho2MM01}) and using
eq. (\ref{SPint}). The exponentials cancel but the prefactor $1/\hd^2$ 
makes the second term in eq. (\ref{rho2MM01}) vanish, leading to the quenched
result. Again these limits  $\hd\to0$ and $\hd\to\infty$ can also be checked
for the gap probability.

As the last and probably most physically relevant 
example we consider the partially quenched case of 
$N_1=0$ and $N_2=2$ flavours with two possibly non-degenerate 
masses $\hm_1$ and $\hm_2$
\be
 E_{S\ 0,0}^{(0+2)}(\hs,0)= \exp\left[-\frac14\hs^2-\hd^2\right] 
\frac{\det\left[
\begin{array}{cc} 
Q_{S}(\hm_1;t=1)& \partial_t Q_{S}(\hm_1;t)\Big|_{t=1}\\
Q_{S}(\hm_2;t=1)& \partial_t Q_{S}(\hm_2;t)\Big|_{t=1}\\
\end{array}
\right]
}{I_0(\hm_1)\hm_2I_1(\hm_2)-I_0(\hm_2)\hm_1 I_1(\hm_1)
}\ .
\label{E2MM2}
\ee
When setting $\hmu_2=0$ to have the sea quarks of flavour $N_2$ 
free of chemical potential we simply have $\hd=-\hmu_1\neq0$. 
The limit of equal
masses can also be taken at the expense of a further derivative within each
determinant. As stressed before, 
this case should be particularly useful for lattice
gauge theory simulations, since it corresponds to ordinary configurations
without chemical potential. 
\begin{figure*}[ht]
  \unitlength1.0cm
  \epsfig{file=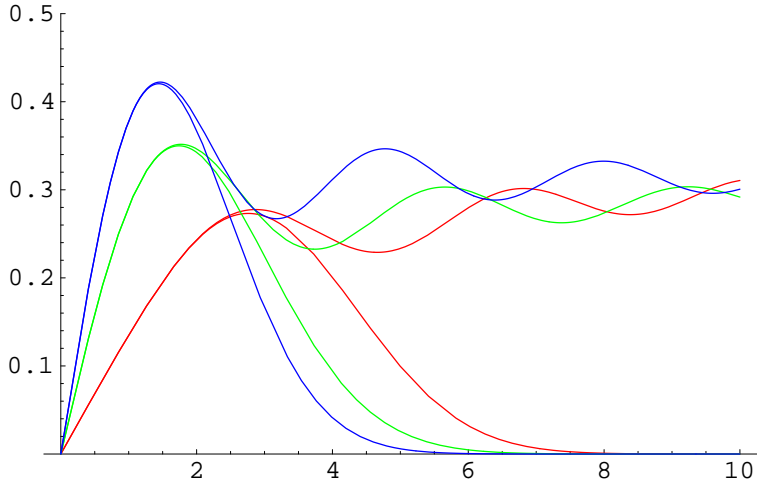,clip=,width=10cm}
  \caption{ 
    \label{fig:5} 
The eigenvalue density and first eigenvalue for $N_f=0+2$ 
with imaginary
    chemical potential $\hd=1$ (low red), 3 (middle green), and 10 (upper
    blue curve), at fixed quark masses of flavour $N_2$
    $\hat{m}_1=3$, $\hat{m}_2=4$.
}
\end{figure*}

The comparison to the spectral density given by\footnote{There is a term
  missing in the bottom right of the $3\times3$ matrix in the relevant
  formula eq. (3.53) of \cite{ADOS}.}
\bea
\rho_{1,0}^{(0+2)}(\hx)&=& \rho^Q_{1M\!M}(\hx)
-\ \exp\left[-\frac12\hd^2\right] \hx\ 
\Big(\hm_1 I_1(\hm_1)I_0(\hm_2)-\hm_2I_0(\hm_1)I_1(\hm_2)\Big)^{-1}
\label{rho2MM02}\\
&\times&\left[
 \int_0^1dt t e^{\frac12\hd^2 t^2}J_0(\hx t)I_0(\hm_1 t)
\Big(-I_0(\hm_2)(\hx J_1(\hx)+\hd^2J_0(\hx)) -                        
\hm_2 I_1(\hm_2)J_0(\hx)\Big)
\right.
\nn\\
&&\left.+
\int_0^1dt t e^{\frac12\hd^2 t^2}J_0(\hx t)I_0(\hm_2 t)
\Big(I_0(\hm_1)(\hx J_1(\hx)+\hd^2J_0(\hx))+
\hm_1I_1(\hm_1)J_0(\hx)\Big)\right],
\nn
\eea
is shown in fig.\ref{fig:5}. The one-matrix model result 
with two flavours is again recovered by setting $\hd=0$
given by eq. (\ref{rho1MMN2}).
For $\hd=1$ the curve is close to the two-flavour
one-matrix result, see fig. \ref{fig:2}, 
and for
$\hd\gg1$ the curves approach the quenched one-matrix quantities,
see fig. \ref{fig:2}.
This matching can once more be checked analytically by taking $\hd\to\infty$ 
\begin{figure*}[h]
  \unitlength1.0cm
  \epsfig{file=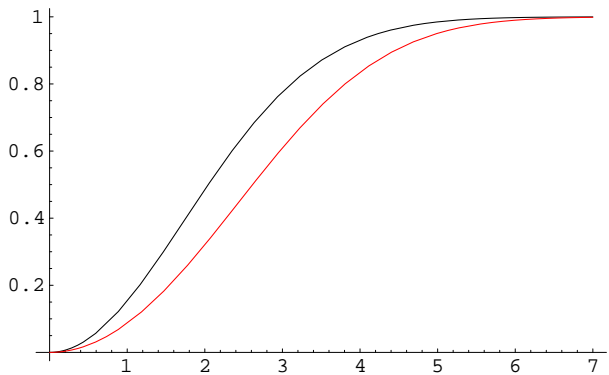,clip=,width=8cm}
  \caption{ 
    \label{fig:6} 
The integrated lowest eigenvalue distribution $1 - E_{0,0}(s)$
for $1+1$ flavours eq. (\ref{E2MM11}) (lower red) 
and $0+2$ flavours eq. (\ref{E2MM2})
(upper 
    black    curve), both at masses  $\hat{m}_1=3$, $\hat{m}_2=4$ and
    $\hd=3$. 
}
\end{figure*}
of eq. (\ref{rho2MM02}). Using eq. (\ref{SPint}) as well as the cancellation
of the two terms proportional to $\hd^2$ in the last two lines of
eq. (\ref{rho2MM02}) leads again to a complete quenching of all flavours. 
The same can be checked for the gap probability.

\begin{figure*}[ht]
  \unitlength1.0cm
  \epsfig{file=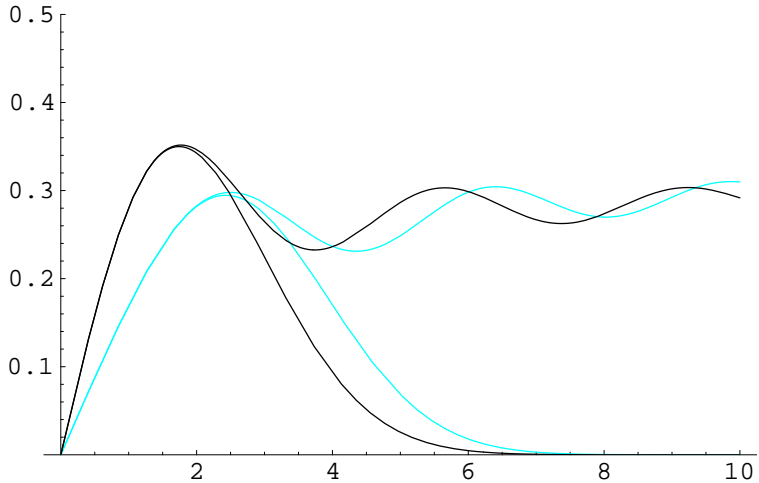,clip=,width=10cm}
  \caption{ 
    \label{fig:7} 
The density and first eigenvalue for $1+1$ flavours (right blue) 
vs. $0+2$ flavours (left 
    black    curve), both at masses  $\hat{m}_1=3$, $\hat{m}_2=4$ and fixed
    $\hd=3$. 
}
\end{figure*}

Finally we can also compare directly the gap probabilities in our two-matrix
theory for $1+1$ flavours and $0+2$ partially quenched 
flavours, where in fig. \ref{fig:6} we show $1 - E_{0,0}(s)$.
For $\hd=1$ the difference is still 
small but it grows rapidly with increasing $\hd$ since both
theories converge towards different limits for $\hd\to\infty$ as was pointed
out earlier.
Equivalently this 
results into the following comparison for the densities and first
eigenvalues shown in fig. \ref{fig:7}. 
It should be noted here that the quantity $1 - E_{0,0}(s)$ is the
integrated lowest eigenvalue distribution, by some considered a
convenient quantity for comparison with the lattice gauge
theory data.

We end this section by pointing out that there is no analogous
computation of a $\mu$-dependent Dirac eigenvalue distribution
in the fully quenched case. The mixed two-point spectral correlation
function has non-trivial $\mu$-dependence \cite{Imiso} (and this
dependence allows for the determination of a quenched value of
$F_{\pi}$ using this technique). But the one-point function
is $\mu$-independent in chiral perturbation theory because it
is generated by the addition of just one valence quark; it 
is $\mu$-independent to all orders in chiral perturbation
theory because the valence pions do not carry net baryon charge.

\sect{Conclusions and outlook}\label{conc}

The two main results of this paper are the following. We have shown how
individual distributions of the lowest-lying eigenvalues of
Dirac operators that are subjected to two different external Abelian
vector potentials (``imaginary chemical potential'') can be 
derived from field theory. The results have been given
in terms of generalised gap probabilities from which the distributions
can all be derived. 

To compute the gap probabilities from field theory
one needs to know spectral correlation functions, all of which can
be given a well-defined meaning in the field theoretical setting. In 
particular, in the scaling region known as the $\epsilon$-regime,
these eigenvalue distributions can be derived from the corresponding
effective theory, the chiral
Lagrangian. To make these computations concrete, we have
used the equivalent Random Two-Matrix Theory to derive the distribution
of the lowest Dirac operator eigenvalue in the $\epsilon$-regime of
QCD with imaginary chemical potential. As stressed in the 
introduction, these analytical formulas may provide a very convenient
way of determining simultaneously 
the infinite-volume chiral condensate $\Sigma$
and the pion decay constant $F_{\pi}$ by means of numerical
simulations in lattice gauge theory. 

We have given explicit formulas for the lowest individual distribution
in terms of a new kernel,
both in the case of full QCD with imaginary chemical potential, and
for the analogue of partially quenched QCD in which quarks are dynamical,
but do not carry chemical potential. Especially the latter
may provide the most useful formulation in terms of comparisons with
numerical lattice data.   

\ \\

{\sc Acknowledgements}:~
This work was supported by 
EPSRC grant EP/D031613/1 (G.A.) and 
EU network ENRAGE MRTN-CT-2004-005616. We would like to thank Francesco Basile
and Leonid Shifrin for discussions.

\begin{appendix}

\sect{A determinant identity}
\label{detid}

In this appendix we prove the following identity for any number of $N_2$
flavours 
\bea
&&\left| 
\begin{array}{ccccc}
\Lh_0(m_1) & \ldots & (1-\tau)^{-k}\Lh_k(m_1) & \ldots & 
(1-\tau)^{-(N+N_2-1)}\Lh_{N+N_2-1}(m_1)\\
\cdots & &\cdots & &\cdots\\
\Lh_0(m_{N_2}) & \ldots & (1-\tau)^{-k}\Lh_k(m_{N_2}) & \ldots & 
(1-\tau)^{-(N+N_2-1)}\Lh_{N+N_2-1}(m_{N_2})\\
\Lh_0(x_1) & \ldots & \Lh_k(x_1) & \ldots & \Lh_{N+N_2-1}(x_1)\\
\cdots & &\cdots & &\cdots\\
\Lh_0(x_N) & \ldots & \Lh_k(x_N) & \ldots & \Lh_{N+N_2-1}(x_N)\\
\end{array}
\right|=
\label{idB}\\
&&\nn\\
=&&\left| 
\begin{array}{ccccc}
\Lh_0(\frac{m_1}{\tau}) & \ldots & \tau^k(1-\tau)^{-k}\Lh_k(\frac{m_1}{\tau}) 
& \ldots & \tau^{N+N_2-1}(1-\tau)^{-(N+N_2-1)}\Lh_{N+N_2-1}(\frac{m_1}{\tau})\\
\cdots & &\cdots & &\cdots\\
\Lh_0(\frac{m_{N_2}}{\tau}) & \ldots 
& \tau^k(1-\tau)^{-k}\Lh_k(\frac{m_{N_2}}{\tau}) 
& \ldots & \tau^{N+N_2-1}(1-\tau)^{-(N+N_2-1)}
\Lh_{N+N_2-1}(\frac{m_{N_2}}{\tau})\\
1& \ldots & x_1^k & \ldots & x_1^{N+N_2-1}\\
\cdots & &\cdots & &\cdots\\
1& \ldots & x_N^k & \ldots & x_N^{N+N_2-1}\\
\end{array}
\right|\ .
\nn
\eea
Here we use the notation $\Lh_n(x)=x^n+\ldots$ for the Laguerre polynomials in
monic normalisation
\be
\Lh_n(x)\equiv (-1)^n n!\ L_n^\nu(x)\ =\ \sum_{j=0}^n (-1)^{n+j}
\frac{n!(n+\nu)!}{(n-j)!(\nu+j)!j!} \ x^j\ \ .
\label{LmonicA}
\ee

For simplicity 
we will prove the identity for one flavour $N_2=1$ first, by induction in $N$.
For $N=1$ we have that 
\be
\left| 
\begin{array}{cc}
\Lh_0(m) & \frac{1}{1-\tau}\Lh_1(m)\\
\Lh_0(x) & \Lh_1(x)\\
\end{array}
\right|
=
\left| 
\begin{array}{cc}
1 & \frac{1}{1-\tau}(m-\nu-1)\\
1 & x-\nu-1\\
\end{array}
\right|
=\left| 
\begin{array}{cc}
1 & \frac{\tau}{1-\tau}(\frac{m}{\tau}-\nu-1)\\
1 & x\\
\end{array}
\right|
=
\left| 
\begin{array}{cc}
\Lh_0(\frac{m}{\tau}) & \frac{1}{1-\tau}\Lh_1(\frac{m}{\tau})\\
1 & x\\
\end{array}
\right|
\ ,
\label{start}
\ee
by adding $\nu+1$ times the first column to the second column.
Next we do the induction step,
 \bea
&&\left| 
\begin{array}{ccccc}
\Lh_0(m) & \ldots & (1-\tau)^{-k}\Lh_k(m) & \ldots & (1-\tau)^{-(N+1)}
\Lh_{N+1}(m)\\
\Lh_0(x_1) & \ldots & \Lh_k(x_1) & \ldots & \Lh_{N+1}(x_1)\\
\cdots & &\cdots & &\cdots\\
\Lh_0(x_{N}) & \ldots & \Lh_k(x_{N}) & \ldots & \Lh_{N+1}(x_{N})\\
\end{array}
\right|=
\label{iduct}\\
&&\nn\\
=&&\left| 
\begin{array}{cccccc}
\Lh_0(\frac{m}{\tau}) & \ldots & \tau^k(1-\tau)^{-k}\Lh_k(\frac{m}{\tau}) 
& \ldots & \tau^N(1-\tau)^{-N}\Lh_N(\frac{m}{\tau}) &
(1-\tau)^{-(N+1)}\Lh_{N+1}({m}) \\
1& \ldots & x_1^k & \ldots & x_1^N &\Lh_{N+1}(x_1)\\
\cdots & &\cdots & &\cdots  &\cdots\\
1& \ldots & x_{N}^k & \ldots & x_{N}^N& \Lh_{N+1}(x_{N})\\
\end{array}
\right|\ .
\nn
\eea
Here we have expanded with respect to the last column and used the induction
assumption for $N$, as well as the fact that the sub-determinant containing
only $x$-variables of monic Laguerre polynomials can be replaced by the
Vandermonde determinant. 

To get monic powers in the last column (except in the first element) we
subsequently subtract multiples of columns from the left, using
eq. (\ref{LmonicA}), and we obtain
\be
\left|
\begin{array}{cccccc}
\Lh_0(\frac{m}{\tau}) & \ldots & \tau^k(1-\tau)^{-k}\Lh_k(\frac{m}{\tau}) 
& \ldots & \tau^N(1-\tau)^{-N}\Lh_N(\frac{m}{\tau}) &
P(m)
\\
1& \ldots & x_1^k & \ldots & x_1^N & x_1^{N+1}\\
\cdots & &\cdots & &\cdots  &\cdots\\
1& \ldots & x_{N}^k & \ldots & x_{N}^N& x_{N}^{N+1}\\
\end{array}
\right|\ .
\ee
The first element in the last column now reads
\be
P(m) \ = \ (1-\tau)^{-(N+1)}\Lh_{N+1}({m}) -\sum_{j=0}^N 
(-1)^{N+1+j}
\frac{(N+1)!(N+1+\nu)!}{(N+1-j)!(\nu+j)!j!}\frac{\tau^j}{(1-\tau)^j}
\Lh_j\left(\frac{m}{\tau}\right) \ .
\label{XL}
\ee
As a last step we need to show that 
$P(m)=\tau^{N+1}(1-\tau)^{-(N+1)}\Lh_{N+1}(\frac{m}{\tau})$. This relation
holds due to the following identity \cite{Wolfram}, which can be easily proven
by induction. It is expressed in terms of usual
non-monic Laguerre polynomials 
\be
L_{N+1}^\nu(m)=\sum_{j=0}^{N+1} 
\frac{(N+1+\nu)!}{(N+1-j)!(\nu+j)!} \tau^j(1-\tau)^{N+1-j}
L_j^\nu\left(\frac{m}{\tau}\right) \ ,
\label{Lid}
\ee
which finishes the first part of our proof. As a remark which is
useful for the main
text this identity is usually quoted as  \cite{Wolfram}
\be
L_{n}^\nu(zw)=\sum_{j=0}^{n} 
\frac{(n+\nu)!}{(n-j)!(\nu+j)!} w^j(1-w)^{n-j}
L_j^\nu\left(z\right) \ .
\label{Lid2}
\ee

In the above it was not essential in the manipulation of columns that we had
one mass flavour $N_2=1$ only. 
We can in fact do an inductive proof in the column
number $k$ for any $N_2$ and $N$,
\bea
&&\left| 
\begin{array}{ccccc}
\Lh_0(m_1) & \ldots & (1-\tau)^{-k}\Lh_k(m_1) & \ldots & 
(1-\tau)^{-(N+N_2-1)}\Lh_{N+N_2-1}(m_1)\\
\cdots & &\cdots & &\cdots\\
\Lh_0(m_{N_2}) & \ldots & (1-\tau)^{-k}\Lh_k(m_{N_2}) & \ldots & 
(1-\tau)^{-(N+N_2-1)}\Lh_{N+N_2-1}(m_{N_2})\\
\Lh_0(x_1) & \ldots & \Lh_k(x_1) & \ldots & \Lh_{N+N_2-1}(x_1)\\
\cdots & &\cdots & &\cdots\\
\Lh_0(x_N) & \ldots & \Lh_k(x_N) & \ldots & \Lh_{N+N_2-1}(x_N)\\
\end{array}
\right|=
\label{idk}\\
&&\nn\\
=&&\left| 
\begin{array}{cccccc}
\Lh_0(\frac{m_1}{\tau}) & \ldots & \frac{\tau^k}{(1-\tau)^{k}}
\Lh_k(\frac{m_1}{\tau}) 
&\frac{1}{(1-\tau)^{k+1}}\Lh_{k+1}(m_{1}) 
& \ldots & \frac{1}{(1-\tau)^{N+N_2-1}}
\Lh_{N+N_2-1}({m_1})\\
\cdots & & \cdots&\cdots & &\cdots\\
\Lh_0(\frac{m_{N_2}}{\tau}) & \ldots 
& \frac{\tau^k}{(1-\tau)^{k}}\Lh_k(\frac{m_{N_2}}{\tau})
& \frac{1}{(1-\tau)^{k+1}}\Lh_{k+1}(m_{N_2}) 
& \ldots & \frac{1}{(1-\tau)^{N+N_2-1}}
\Lh_{N+N_2-1}({m_{N_2}})\\
1& \ldots & x_1^k & \Lh_{k+1}(x_1)
&\ldots & \Lh_{N+N_2-1}(x_1)\\
\cdots & &\cdots  &\cdots & &\cdots\\
1& \ldots & x_N^k & \Lh_{k+1}(x_N)&
\ldots &\Lh_{N+N_2-1}(x_N)\\
\end{array}
\right|\ .
\nn
\eea
The start for $k=1$ is trivially true in analogy to eq. (\ref{start}).
The induction step from $k$ to $k+1$ easily follows by subtracting the
left columns for $l\leq k$
from column $k+1$, and using again eq. (\ref{Lid}) for $N+1\to k+1$.
Putting $k=N+N_2$ ends the proof.


\sect{An identity for Laguerre polynomials}
\label{B}

The relation we show here is given in eq. (\ref{LidII}),
\be
L_{m}^{-1}(x)\ =\ 
\sum_{j=0}^{m}(-)^j{j+k \choose k} L_{m-j}^{j+k}(x) \ , 
\label{LidIIB}
\ee
where the right hand side is independent of $k$. It 
follows from a known identity eq. (4.4.1.14) in \cite{Prudnikov} 
\be
\frac{1}{(\beta)_m} L_m^{\al+\beta-1}(x)\ =\
\sum_{i=0}^m\frac{1}{(m-i)!(\beta)_i}L_i^{\al-i}(x)  \ .
\ee
Here $(\beta)_m$ is the Pochhammer symbol. 
In choosing $\al=-\beta$ we obtain
\bea
\frac{(-)^m(\al-m)!}{\al!} L_m^{-1}(x) &=&
\sum_{i=0}^m\frac{(-)^{i}(\al-i)!}{(m-i)!\al!}L_i^{\al-i}(x)  \ .\nn\\
&=&  \sum_{j=0}^m\frac{(-)^{m-j}(\al-m+j)!}{j!\al!}L_{m-j}^{\al-m+j}(x)\ .
\eea
In the second step we have changed summation from $i$ to $j=m-i$. Finally
choosing $\al=m+k$ we obtain eq. (\ref{LidIIB}) above. 

\end{appendix}


\begin{thebibliography}{X}
\bibitem{LS}
J.~Gasser and H.~Leutwyler,
  Phys.\ Lett.\  B {\bf 184} (1987) 83;
H.~Neuberger,
  Phys.\ Rev.\ Lett.\  {\bf 60} (1988) 889;
H.~Leutwyler and A.~Smilga,
  Phys.\ Rev.\  D {\bf 46} (1992) 5607.

\bibitem{Jacetal}
E.~V.~Shuryak and J.~J.~M.~Verbaarschot,
  Nucl.\ Phys.\  A {\bf 560} (1993) 306
  [hep-th/9212088];
J.~J.~M.~Verbaarschot and I.~Zahed,
  Phys.\ Rev.\ Lett.\  {\bf 70} (1993) 3852
  [hep-th/9303012];
J.~J.~M.~Verbaarschot,
  Phys.\ Rev.\ Lett.\  {\bf 72} (1994) 2531
  [hep-th/9401059];
G.~Akemann, P.~H.~Damgaard, U.~Magnea and S.~Nishigaki,
  Nucl.\ Phys.\  B {\bf 487} (1997) 721
  [hep-th/9609174].

\bibitem{DOTV}
P.~H.~Damgaard, J.~C.~Osborn, D.~Toublan and J.~J.~M.~Verbaarschot,
  Nucl.\ Phys.\  B {\bf 547} (1999) 305
  [hep-th/9811212].

\bibitem{BA}
F.~Basile and G.~Akemann,
JHEP {\bf 12} (2007) 043
[arXiv:0710.0376v2 [hep-th]].

\bibitem{AD}
G.~Akemann and P.~H.~Damgaard,
  Phys.\ Lett.\  B {\bf 583} (2004) 199
  [hep-th/0311171].

\bibitem{Sigmasim0}
M.~E.~Berbenni-Bitsch, A.~D.~Jackson, S.~Meyer, A.~Sch\"afer, 
J.~J.~M.~Verbaarschot and T.~Wettig,
  Nucl.\ Phys.\ Proc.\ Suppl.\  {\bf 63} (1998) 820
  [hep-lat/9709102];
P.~H.~Damgaard, U.~M.~Heller, R.~Niclasen and K.~Rummukainen,
  Phys.\ Lett.\  B {\bf 495} (2000) 263
  [hep-lat/0007041].

\bibitem{Sigmasim}
R.~G.~Edwards, U.~M.~Heller, J.~E.~Kiskis and R.~Narayanan,
  Phys.\ Rev.\ Lett.\  {\bf 82} (1999) 4188
  [arXiv:hep-th/9902117];
L.~Giusti, M.~L\"uscher, P.~Weisz and H.~Wittig,
  JHEP {\bf 0311} (2003) 023
  [hep-lat/0309189];
T.~A.~DeGrand and S.~Schaefer,
  Phys.\ Rev.\  D {\bf 72} (2005) 054503
  [hep-lat/0506021];
J.~Wennekers and H.~Wittig,
  JHEP {\bf 0509} (2005) 059
  [arXiv:hep-lat/0507026];
T.~DeGrand, R.~Hoffmann, S.~Schaefer and Z.~Liu,
  Phys.\ Rev.\  D {\bf 74} (2006) 054501
  [arXiv:hep-th/0605147];
H.~Fukaya {\it et al.}  [JLQCD Collaboration],
  Phys.\ Rev.\ Lett.\  {\bf 98} (2007) 172001
  [arXiv:hep-lat/0702003].

\bibitem{Imiso}
P.~H.~Damgaard, U.~M.~Heller, K.~Splittorff and B.~Svetitsky,
  Phys.\ Rev.\  D {\bf 72} (2005) 091501
  [hep-lat/0508029];
P.~H.~Damgaard, U.~M.~Heller, K.~Splittorff, B.~Svetitsky and D.~Toublan,
  Phys.\ Rev.\  D {\bf 73} (2006) 074023
  [hep-lat/0602030];
  Phys.\ Rev.\  D {\bf 73} (2006) 105016
  [hep-th/0604054].

\bibitem{MT}
T.~Mehen and B.~C.~Tiburzi,
  Phys.\ Rev.\  D {\bf 72} (2005) 014501
  [hep-lat/0505014].

\bibitem{Luz}
M.~Luz,
  Phys.\ Lett.\  B {\bf 643} (2006) 235
  [hep-lat/0607022].

\bibitem{AW}
J.~C.~Osborn and T.~Wettig,
  PoS {\bf LAT2005} (2006) 200
  [hep-lat/0510115];
 G.~Akemann and T.~Wettig,
  Phys.\ Rev.\ Lett.\  {\bf 92} (2004) 102002,
  Erratum-ibid.\  {\bf 96} (2006) 029902
  [hep-lat/0308003];
G. Akemann, J. Bloch, L. Shifrin and T. Wettig,
Phys. Rev. Lett. {\bf 100} (2008) 032002
[arXiv:0710.2865v2 [hep-lat]].

\bibitem{ADOS}
G.~Akemann, P.~H.~Damgaard, J.~C.~Osborn and K.~Splittorff,
Nucl. Phys. {\bf B766} (2007) 34 
[hep-th/0609059].

\bibitem{Tom}
P.~H.~Damgaard, T.~DeGrand and H.~Fukaya,
  JHEP {\bf 0712} (2007) 060
  [arXiv:0711.0167 [hep-lat]].

\bibitem{VA}
P.~H.~Damgaard, P.~Hernandez, K.~Jansen, M.~Laine and L.~Lellouch,
  Nucl.\ Phys.\  B {\bf 656} (2003) 226
  [arXiv:hep-lat/0211020];
L.~Giusti, P.~Hernandez, M.~Laine, P.~Weisz and H.~Wittig,
  JHEP {\bf 0401} (2004) 003
  [arXiv:hep-lat/0312012].

\bibitem{Necco}
S.~Necco,
  PoS {\bf LAT2007} (2007) 021 [arXiv:0710.2444 [hep-lat]].

\bibitem{AD07}
G.~Akemann and P.~H.~Damgaard,
PoS {\bf LAT2007} (2007) 166 
[arXiv:0709.0484 [hep-lat]].

\bibitem{Tom0}
T.~DeGrand and S.~Schaefer,
PoS {\bf LAT2007} (2007) 069  [arXiv:0709.2889 [hep-lat]].

\bibitem{A07}
G.~Akemann, 
Acta Phys. Pol. B. {\bf 38} (2007) 3981
[arXiv:0710.2905v1 [hep-th]].

\bibitem{AV}
G.~Akemann and G.~Vernizzi,
  Nucl.\ Phys.\ B. {\bf 660} (2003) 532
  [hep-th/0212051].

\bibitem{DNW}
  S.~M.~Nishigaki, P.~H.~Damgaard and T.~Wettig,
  Phys.\ Rev.\ D. {\bf 58} (1998) 087704
  [hep-th/9803007];
P.~H.~Damgaard and S.~M.~Nishigaki,
  Phys.\ Rev.\ D. {\bf 63} (2001) 045012
  [hep-th/0006111].

\bibitem{Prudnikov} A.P. Prudnikov, Yu.A. Brychkov and O.I. Marichev, 
{\it Integrals and Series Vol.2, Special Functions},
New York, Gordon and Breach Science Publishers, 1986.

\bibitem{DN}
  P.~H.~Damgaard and S.~M.~Nishigaki,
  Nucl.\ Phys.\  B {\bf 518} (1998) 495
  [arXiv:hep-th/9711023].

\bibitem{Wolfram}
Wolfram Web Resources, 
http://functions.wolfram.com/05.08.23.0004.01


\end{thebibliography}
\end{document}